\documentclass[12pt]{article}
\usepackage{amsmath}
\usepackage{amsfonts}
\usepackage{amsthm}
\usepackage{graphicx}
\newtheorem{thm}{Theorem}[section]
\newtheorem{pro}[thm]{Proposition}
\newtheorem{cor}[thm]{Corollary}

\newtheorem{rem}[thm]{Remark}
\newtheorem{defn}[thm]{Definition}

\begin{document}

\title{Solutions of Non-Integrable Equations by the Hirota Direct Method}

\vspace{5cm}

\author{ Asl{\i} Pekcan\\
{\small  Department of Mathematics, Faculty of Sciences}\\
{\small Bilkent University, 06800 Ankara, Turkey}\\}

\maketitle

\begin{abstract}
We show that we can also apply the Hirota method to some
non-integrable equations. For this purpose, we consider the
extensions of the Kadomtsev-Petviashvili (KP) and the Boussinesq
(Bo) equations. We present several solutions of these equations.
\end{abstract}

\vspace{10mm}

\section{Introduction} The Hirota direct method is one of the famous method
to construct multi-soliton solutions of integrable nonlinear
partial differential equations. Hirota gave the exact solution of
the Korteweg-de Vries (KdV) equation  for multiple collisions of
solitons by using the Hirota direct method in $1971$
\cite{HirotaKdV}. In his successive articles, he dealt also with
many other nonlinear evolution equations such as the modified
Korteweg-de Vries (mKdV) \cite{HirotamKdV}, sine-Gordon (sG)
\cite{HirotasG}, nonlinear Schr\"{o}dinger (nlS) \cite{HirotanlS}
and Toda lattice (Tl) \cite{HirotaTl} equations.

The first step of this method is to transform the nonlinear
partial differential or difference equation into a quadratic form
in dependent variables. The new form of the equation is called
'bilinear form'. In the second step, we write the bilinear form
the equation as a polynomial of a special differential operator,
Hirota D-operator. This polynomial of D-operator is called 'Hirota
bilinear form'. In fact, some equations may not be written in the
Hirota bilinear form but perhaps in trilinear or multilinear forms
\cite{GrammaticosRH}. The last step of the method is using the
finite perturbation expansion in the Hirota bilinear form. We
analyze the coefficients of the perturbation parameter and its
powers separately. Here the information we gain makes us to reach
the exact solution of the equation.

The equations having Hirota bilinear form possesses automatically
one- and two-soliton solutions. But when we try to construct the
three-soliton solutions we come across a very restrictive
condition. This condition was used as a powerful tool to search
the integrability of the equations by Hietarinta
\cite{Hietarinta3ss}. Hietarinta also used this condition to
produce new integrable equations in his articles \cite{HKdVtype},
\cite{HMKdVtype}, \cite{HSGtype}, \cite{HNLStype}.

Most of the works dealt with the Hirota direct method is about the
integrable equations. But in this work, we show that the Hirota
direct method also can be used to find exact solutions of some
non-integrable nonlinear partial differential equations. For
illustration we consider an extension of the
Kadomtsev-Petviashvili (KP) equation,
\begin{equation}
(u_t-6uu_x+u_{xxx})_x+3u_{yy}+au_{tt}+bu_{ty}+c\nabla^2 u=0
\end{equation}
where $a$, $b$ and $c$ are constants and $\nabla^2
u=u_{x_1x_1}+u_{x_2x_2}+...+u_{x_mx_m}$ for $x_j$, $j=1,2,...,m$
are independent variables. This extension of the KP equation has
bilinear and Hirota bilinear form so the Hirota direct method is
applicable. But when we obtain exact solutions, we shall consider
the case $c=0$. In this case the equation turns out to be
\begin{equation}
(u_t-6uu_x+u_{xxx})_x+3u_{yy}+au_{tt}+bu_{ty}=0
\end{equation}
and we call it as the extended Kadomtsev-Petviashvili (eKP)
equation. This equation is integrable if $a=b^2/12$. The equation
with this condition is equivalent to the KP equation. We find
exact solution of this equation by using the Hirota method for all
$a$, $b$. Another example to this fact is the extension of the
Boussinesq (Bo) equation which is
\begin{equation}
u_{tt}-u_{xx}+3(u^2)_{xx}-u_{xxxx}+au_{yy}+bu_{ty}+c\nabla^2 u=0.
\end{equation}
Similar to the extension of the KP equation, $a$, $b$ and $c$ are
constants. When $c\neq 0$, we can also apply the Hirota method to
this equation since it has bilinear and Hirota bilinear form. But
again, when we consider the exact solutions, we shall take $c=0$.
So we deal with the equation
\begin{equation}
u_{tt}-u_{xx}+3(u^2)_{xx}-u_{xxxx}+au_{yy}+bu_{ty}=0
\end{equation}
which we call the extended Boussinesq (eBo) equation. The eBo
equation is integrable if $a=b^2/4$. Indeed, under this condition,
it is equivalent to the Bo equation. The Hirota method gives the
exact solutions of the eBo equation for all $a$, $b$. Before
passing to the application of the Hirota direct method, let us see
how this method works.

\subsection{The Hirota Direct Method}

We review the Hirota direct method in four steps by following
Hietarinta's article \cite{Hietarinta} closely. Let
$F[u]=F(u,u_x,u_t,...)$ be a nonlinear partial differential
equation.

\vspace{2mm}

\noindent \textbf{Step $1$}: \textit{Bilinearization}: We
transform $F[u]$ to a quadratic form in the dependent variables by
a bilinearizing transformation $u = T[f(x,t,...),g(x,t,...)]$. We
call this form the bilinear form of $F[u]$. Note that for some
equations we may not find such a transformation.

\vspace{2mm}

\noindent \textbf{Step $2$}: \textit{Transformation to the Hirota
bilinear form}:
\begin{defn}
 Let $S: {\mathbb{C}}^n \rightarrow {\mathbb{C}}$ be a space of
differentiable functions. Then Hirota D-operator $D:S\times S
\rightarrow S$ is defined as
\begin{multline}
[D_x^{m_1}D_t^{m_2}...]\{f.g
\}=[(\partial_x-\partial_{x'})^{m_1}(\partial_t-\partial_{t'})^{m_2}...]
f(x,t,...)\\ \times g(x',t',...)|_{x'=x,t'=t,...}
\end{multline}
where $m_i$, $i=1,2,...$ are positive integers and $x,t,...$ are
independent variables.
\end{defn}
By using some sort of combination of Hirota D-operator, we try to
write the bilinear form of $F[u]$ as a polynomial of D-operator,
say $P(D)$. Let us state some propositions and corollaries on
$P(D)$  \cite{Hietarinta}.\\
\begin{pro} Let $P(D)$ act on two differentiable functions $f(x,t,...)$
and $g(x,t,...)$. Then we have
\begin{equation}
P(D)\{f.g \} = P(-D)\{g.f \}.
\end{equation}
\end{pro}
\begin{cor}Let $P(D)$ act on two differentiable functions $f(x,t,...)$
and $g=1$, then we have
\begin{equation}
P(D)\{f.1 \} = P(\partial)f \quad , \quad P(D)\{1.f \} =
P(-\partial)f.
\end{equation}
\end{cor}
\begin{pro} Let P(D) act on two exponential functions $e^{\theta_1}$ and
$e^{\theta_2}$ where $\theta_i = k_ix +w_it+...+\alpha_i$ with
$k_i,w_i,...,\alpha_i$ are constants for $i=1, 2$.  Then we have
\begin{equation}
 P(D)\{e^{\theta_1}.e^{\theta_2}\} =
P(k_1-k_2,...,\alpha_1-\alpha_2)e^{\theta_1+\theta_2}.
\end{equation}
\end{pro}
\noindent For a shorter notation, we use $P(p_1-p_2)$ instead of
$P(k_1-k_2,...,\alpha_1-\alpha_2)$.
\begin{cor}If we have
$P(D)\{a.a \}=0$ where $a$ is any nonzero constant then we have
$P(0,0,...)=0$. \end{cor}
\begin{defn}
We say that a nonlinear partial differential equation can be
written in Hirota bilinear form if it is equivalent to
\begin{equation}
\sum_{\alpha,\beta=1}^{m}P_{\alpha\beta}^{\eta}(D)f^\alpha .
f^\beta = 0, \quad \eta=1, ..., r
\end{equation}
for some m,r and linear operators $P_{\alpha\beta}^{\eta}(D)$. The
$f^i$'s are new dependent variables.
\end{defn}
\begin{rem}
 There is no systematic way
to write a nonlinear partial differential equation in Hirota
bilinear form.
\end{rem}
\begin{rem} For some nonlinear partial differential equations we may
need more than one Hirota bilinear equation.
\end{rem}

\vspace{2mm}

\noindent \textbf{Step $3$}: \textit{Application of the Hirota
perturbation}: We substitute the finite perturbation expansions of
the differentiable functions $f(x,t,...)$ and $g(x,t,...)$ which
are
\begin{equation}
f(x,t,...) = f_0 + \sum_{m=1}^{N}\varepsilon^mf_m(x,t,...) \quad ,
\quad g(x,t,...) = g_0 + \sum_{m=1}^{N}\varepsilon^mg_m(x,t,...)
\end{equation}
into the Hirota bilinear form. Here $f_0$, $g_0$ are constants
with the condition $(f_0,g_0)\neq(0,0)$ to avoid the trivial
solution. For the sake of applicability of the method we take the
functions $f_m$ and $g_m$, $m=1,...,N$ as exponential functions.
$\varepsilon$ is a constant called the perturbation parameter. For
instance for $N=2$, we take
\begin{equation}
f=f_0+\varepsilon f_1+\varepsilon^2 f_2 \quad , \quad
g=g_0+\varepsilon g_1+ \varepsilon^2 g_2
\end{equation}
where $f_1=e^{\theta_1}+e^{\theta_2}$ for
$\theta_i=k_ix+\omega_it+...+\alpha_i$, $i=1,2$. We decide what
the other terms of the functions $f$ and $g$ in the process of the
method.

\vspace{2mm}

\noindent \textbf{Step $4$}: \textit{Examination of the
coefficients of the perturbation parameter $\varepsilon$}: We make
the coefficients of the perturbation parameter $\varepsilon$ and
its powers appeared in the Hirota perturbation to vanish. From
these coefficients we obtain the functions $f(x,t,...)$ and
$g(x,t,...)$. Hence by using them in the bilinearizing
transformation $u=T[f(x,t,...),g(x,t,...)]$, we find the exact
solution of $F[u]$.

\vspace{4mm}

\section{Applications of the Hirota Direct Method}

\vspace{4mm}

\subsection{The Extended Kadomtsev-Petviashvili (EKP) \\
Equation }

The extended Kadomtsev-Pethviashvili (eKP) equation is given by
\begin{equation}
(u_t-6uu_x+u_{xxx})_x+3u_{yy}+au_{tt}+bu_{ty}+c\nabla^2 u=0
\end{equation}
which is constructed by adding the terms $au_{tt}$, $bu_{ty}$ and
$\nabla^2 u=u_{x_1x_1}+u_{x_2x_2}+...+u_{x_mx_m}$ multiplied by
$c$ to the Kadomtsev-Petviashvili (KP) equation where $a$, $b$ and
$c$ are constants and $x_j$, $j=1,2,...,m$ are independent
variables. Now let us apply the Hirota direct method to the eKP
equation.

\noindent \textbf{Step $1$}. \textit{Bilinearization}: We use the
bilinearizing transformation
\begin{equation}
u(x,t,y)=-2\partial_x^2 \log f
\end{equation}
 so the bilinear form of eKP is
\begin{multline}\label{eKPbilinear}
f_{tx}f-f_tf_x+f_{xxxx}f-4f_xf_{xxx}+3f_{xx}^2+3f_{yy}f-3f_y^2\\+
aff_{tt}-af_t^2+bf_{ty}f-bf_tf_y+c\sum_{j=1}^m(f_{x_jx_j}f-f_{x_j}^2)
=0.
\end{multline}
\noindent \textbf{Step $2$}. \textit{Transformation to the Hirota
bilinear form}: The Hirota bilinear form of eKP is
\begin{equation}\label{eKPhirotabilinear}
P(D)\{f.f
\}=(D_tD_x+D_x^4+3D_y^2+aD_t^2+bD_tD_y+c\sum_{j=1}^mD_{x_j}^2)\{f.f
\}=0.
\end{equation}

\noindent \textbf{Step $3$.} \textit{Application of the Hirota
perturbation}: Insert $f=1+\sum_{n=1}^N \varepsilon^n f_n$ into
the equation (\ref{eKPhirotabilinear}) so we have
\begin{multline}\label{EKPPERTURBATION}
P(D)\{f.f \}=P(D)\{1.1 \} + \varepsilon P(D)\{f_1.1 +1.f_1
\}\\+...+\varepsilon^{2N} P(D)\{f_N.f_N \}=0.
\end{multline}

\noindent \textbf{Step $4$}: \textit{Examination of the
coefficients of the perturbation parameter $\varepsilon$}: We make
the coefficients of $\varepsilon^m$, $m=1,2,...,N$ appeared in
\ref{EKPPERTURBATION} to vanish. Here we shall consider only the
case $N=3$ and $N=4$. Note that since eKP is not integrable except
if $a=b^2/12$, we call the solutions obtained using the Hirota
method as the $N$-Hirota solution of eKP.

\subsubsection{$N=3$, Three-Hirota Solution of EKP}
 \noindent Here we apply the Hirota direct method by using the
 anzats which is used to construct three-soliton solutions.
 We take $ f = 1 + \varepsilon f_1 + \varepsilon^2 f_2 +
\varepsilon^3 f_3$ where $f_1 = e^{\theta_1}+e^{\theta_2} +
e^{\theta_3}$ with $\theta_i = k_ix + \omega_it + l_iy +
\sum_{j=1}^m r_{ij}x_j+\alpha_i$ for $i = 1,2,3$ and insert it
into (\ref{EKPPERTURBATION}). The coefficient of $\varepsilon^0$
is identically zero since
\begin{equation}
P(D)\{1.1 \}=0.
\end{equation}
By the coefficient of $\varepsilon^1$
\begin{equation}
P(D)\{1.f_1 + f_1.1 \}
=2P(\partial)\{e^{\theta_1}+e^{\theta_2}+e^{\theta_3}\}=0
\end{equation}
we have the relation
\begin{equation}\label{eKPdispersion}
P(p_i)=k_i\omega_i+k_i^4+3l_i^2+a\omega_i^2
+b\omega_il_i+c\sum_j^m r_{ij}^2=0
\end{equation}
for $i=1,2,3$. This relation is called as the dispersion relation.
 Note that when $c$, the coefficient
of $\nabla^2 u$ is not zero, we can apply the Hirota direct
method. But for simplicity, we take $c=0$ in the rest of the
calculations. In this case $\theta_i$ turn out to be $\theta_i =
k_ix + \omega_it + l_iy + \alpha_i$, $i=1,2,3$. From the
coefficient of $\varepsilon^2$ we get
\begin{multline}\label{eKP3ssf_2}
-P(\partial)f_2=\sum_{i<j}^{(3)}[(k_i-k_j)(\omega_i-\omega_j)+(k_i-k_j)^4+3(l_i-l_j)^2\\+a(\omega_i-\omega_j)^2
+b(\omega_i-\omega_j)(l_i-l_j)]e^{\theta_i+\theta_j}
\end{multline}
\noindent where $(3)$ indicates the summation of all possible
combinations of the three elements with
 $i<j$. Thus $f_2$ should be of the form
\begin{equation}
f_2 = A(1,2)e^{{\theta_1}+{\theta_2}} +
A(1,3)e^{{\theta_1}+{\theta_3}} + A(2,3)e^{{\theta_2}+{\theta_3}}
\end{equation}
to satisfy the equation. We insert $f_2$ into the equation
(\ref{eKP3ssf_2}) so we get $A(i,j)$ as {{\small
\begin{equation}
\begin{split}
 A(i,j)&=-\frac{P(p_i-p_j)}{P(p_i+P_j)}\\
 &=\displaystyle \frac{b(\omega_il_j+\omega_jl_i)+2a
\omega_i\omega_j+k_i(\omega_j+4k_j^3)+k_j(\omega_i+4k_i^3)-6k_i^2k_j^2+6l_il_j}
{b(\omega_il_j+\omega_jl_i)+2a
\omega_i\omega_j+k_i(\omega_j+4k_j^3)+k_j(\omega_i+4k_i^3)+6k_i^2k_j^2+6l_il_j}
\end{split}
\end{equation}}}
\noindent where $i,j=1,2,3$, $i<j$. From the coefficient of
$\varepsilon^3$ we obtain
\begin{multline}
P(\partial)\{f_3 \} = -[A(1,2)P(p_3-p_2-p_1)+A(1,3)P(p_2-p_1-p_3)\\
+A(2,3)P(p_1-p_2-p_3)]e^{{\theta_1}+{\theta_2} + {\theta_3}}.
\end{multline}
Hence $f_3$ is in the form $f_3 =Be^{{\theta_1}+{\theta_2} +
{\theta_3}}$ where $B$ is found as {{\small
\begin{equation}\label{eKPBuzun}
B =\displaystyle - \frac{A(1,2)P(p_3-p_1-p_2) +
A(1,3)P(p_2-p_1-p_3) + A(2,3)P(p_1-p_2-p_3)}{P(p_1+p_2+p_3)}.
\end{equation}}}
\noindent The coefficient of $\varepsilon^4$ gives us
\begin{multline}
e^{{2\theta_1}+{\theta_2}+{\theta_3}}
[BP(p_2+p_3)+A(1,2)A(1,3)P(p_2-p_3)]\\+
e^{{\theta_1}+{2\theta_2}+{\theta_3}}
[BP(p_1+p_3)+A(1,2)A(2,3)P(p_1-p_3)]\\+
e^{{\theta_1}+{\theta_2}+{2\theta_3}}
[BP(p_1+p_2)+A(1,3)A(2,3)P(p_1-p_2)]=0
\end{multline}
which is satisfied when
\begin{equation}\label{eKPBkisa}
B = A(1,2)A(1,3)A(2,3).
\end{equation}
To be consistent the two expressions (\ref{eKPBuzun}) and
(\ref{eKPBkisa}) should be equivalent i.e.
\begin{equation}\label{eKP3sscondition}
\begin{split}
B =&\displaystyle - \frac{A(1,2)P(p_3-p_1-p_2) +
A(1,3)P(p_2-p_1-p_3) + A(2,3)P(p_1-p_2-p_3)}{P(p_1+p_2+p_3)}\\
=&A(1,2)A(1,3)A(2,3).
\end{split}
\end{equation}
The above equivalence is satisfied when the condition
\begin{equation}
\begin{split}
&P(p_1-p_2)P(p_1-p_3)P(p_2-p_3)P(p_1+p_2+p_3)\\
+&P(p_1-p_2)P(p_1+p_3)P(p_2+p_3)P(p_3-p_1-p_2)\\
+&P(p_1-p_3)P(p_1+p_2)P(p_2+p_3)P(p_2-p_1-p_3)\\
+&P(p_2-p_3)P(p_1+p_2)P(p_1+p_3)P(p_1-p_2-p_3)=0
\end{split}
\end{equation}
holds. This condition which we call three-Hirota solution
condition $(3HC)$ can also be written as
\begin{equation}
\sum_{\sigma_r=\pm
1}P(\sigma_1p_1+\sigma_2p_2+\sigma_3p_3)P(\sigma_1p_1-\sigma_2p_2)P(\sigma_2p_2-\sigma_3p_3)P(\sigma_1p_1-\sigma_3p_3)=0,
\end{equation}
\noindent $r=1,2,3$. After some simplifications $(3HC)$ turns out
to be
\begin{equation}\label{eKP3HC}
\begin{split}
&(12a-b^2)k_1^2k_2^2k_3^2\Big[2k_1^2w_2w_3l_2l_3+2k_2^2w_1w_3l_1l_3+2k_3^2w_1w_2l_1l_2
\\&+2k_2k_3w_2w_3l_1^2
+2k_1k_3w_1w_3l_2^2+2k_1k_2w_1w_2l_3^2+2k_2k_3w_1^2l_2l_3
\\&+2k_1k_3w_2^2l_1l_3 +2k_1k_2w_3^2l_1l_2
-2k_1k_3w_2w_3l_1l_2-2k_1k_3w_1w_2l_2l_3
\\&-2k_1k_2w_2w_3l_1l_3-2k_1k_2w_1w_3l_2l_3
-2k_2k_3w_1w_3l_1l_2-2k_2k_3w_1w_2l_1l_3
\\&-k_1^2w_3^2l_2^2-k_1^2w_2^2l_3^2-k_2^2w_1^2l_3^2
-k_2^2w_3^2l_1^2-k_3^2w_2^2l_1^2-k_3^2w_1^2l_2^2\Big]=0.
\end{split}
\end{equation}
As we see this condition satisfied when $a=b^2/12$ or for some
relations on $k_i$, $w_i$ and $l_i$ which violate the solitonic
property of the solution. The coefficients of $\varepsilon^5$ and
$\varepsilon^6$ vanish trivially. Let us focus on the condition
(\ref{eKP3HC}). When the relation $a=b^2/12$ holds, the eKP
equation is integrable. In fact, it is transformable to the KP
equation by the transformation
\begin{center}
$u'=u$, $x'=x$, $y'=y$, $t'=t+ \rho y$,
\end{center}
\noindent where $\rho=-b/6=\sqrt{a/3}$. If $a\neq b^2/12$, eKP is
not integrable. In this case, there are other relations which
provide (\ref{eKP3HC}) satisfied. Some of
them are;\\

\noindent \textbf{Case 1.} Any $k_i=0$, $i=1,2,3$, the rest are
different,\\

\noindent \textbf{Case 2.} $k_i=\omega_i$, $i=1,2,3$,\\

\noindent \textbf{Case 3.} $k_i=l_i$, $i=1,2,3$,\\

\noindent \textbf{Case 4.} $\omega_i=l_i$, $i=1,2,3$.\\

\noindent By using any of these cases, we obtain the exact
solutions of eKP.

\subsubsection{$N=4$, Four-Hirota Solution of EKP}
 \noindent Here we apply the Hirota direct method by using the
 anzats which is used to construct four-soliton solutions. We
 take $ f = 1 + \varepsilon f_1 + \varepsilon^2 f_2 +
\varepsilon^3 f_3+\varepsilon^4 f_4$ where $f_1 =
e^{\theta_1}+e^{\theta_2} + e^{\theta_3}+e^{\theta_4}$ with
$\theta_i = k_ix + \omega_it + l_iy + \alpha_i$ for $i = 1,2,3,4$
and insert it into (\ref{EKPPERTURBATION}). We will only consider
the coefficients of $\varepsilon^m$, $m=1, 2, 3, 4, 5$ since the
others vanish identically. By the coefficient of $\varepsilon^1$
\begin{equation}
P(D)\{1.f_1 + f_1.1 \}
=2P(\partial)\{e^{\theta_1}+e^{\theta_2}+e^{\theta_3}+e^{\theta_4}
\}=0
\end{equation} we have the dispersion relation
\begin{equation}
P(p_i)=k_i\omega_i+k_i^4+3l_i^2+a\omega_i^2 +b\omega_il_i=0
\end{equation}
for $i=1,2,3,4$. From the coefficient of $\varepsilon^2$ we get
\begin{multline}
-P(\partial)f_2=P(p_1-p_2)e^{\theta_1+\theta_2}+P(p_1-p_3)e^{\theta_1+\theta_3}+P(p_1-p_4)e^{\theta_1+\theta_4}\\
+P(p_2-p_3)e^{\theta_2+\theta_3}+P(p_2-p_4)e^{\theta_2+\theta_4}+P(p_3-p_4)e^{\theta_3+\theta_4}.
\end{multline}
\noindent Thus $f_2$ should be of the form
\begin{multline}
f_2 = A(1,2)e^{{\theta_1}+{\theta_2}} +
A(1,3)e^{{\theta_1}+{\theta_3}} +A(1,4)e^{\theta_1+\theta_4}\\
+
A(2,3)e^{{\theta_2}+{\theta_3}}+A(2,4)e^{\theta_2+\theta_4}+A(3,4)e^{\theta_3+\theta_4},
\end{multline}
where $A(i,j)$, $i,j=1,2,3,4$ with $i<j$ are obtained as
\begin{equation}
\begin{split}
 A(i,j)&=-\frac{P(p_i-p_j)}{P(p_i+P_j)}\\
 &=\displaystyle \frac{\beta (\omega_il_j+\omega_jl_i)+2\gamma
\omega_i\omega_j+k_i(\omega_j+4k_j^3)+k_j(\omega_i+4k_i^3)-6k_i^2k_j^2+6l_il_j}
{\beta (\omega_il_j+\omega_jl_i)+2\gamma
\omega_i\omega_j+k_i(\omega_j+4k_j^3)+k_j(\omega_i+4k_i^3)+6k_i^2k_j^2+6l_il_j}.
\end{split}
\end{equation}
\noindent After some simplifications, the coefficient of
$\varepsilon^3$ gives
\begin{multline}\label{4ssf_3}
-P(\partial)f_3
=\sum_{i<j<m}^{(4)}[A(i,j)P(p_m-p_i-p_j)+A(i,m)P(p_j-p_i-p_m)\\
+A(j,m)P(p_i-p_j-p_m)]e^{\theta_i+\theta_j+\theta_m}
\end{multline}
\noindent where $(4)$ indicates the summation of all possible
combinations of the four elements with
 $i<j<m$. Hence $f_3$ is of the form
\begin{multline}
f_3 =B(1,2,3)e^{{\theta_1}+{\theta_2} +
{\theta_3}}+B(1,2,4)e^{{\theta_1}+{\theta_2} +
{\theta_4}}\\+B(1,3,4)e^{{\theta_1}+{\theta_3} +
{\theta_4}}+B(2,3,4)e^{{\theta_2}+{\theta_3} + {\theta_4}}.
\end{multline}
We insert $f_3$ into (\ref{4ssf_3}) and obtain {{\small
\begin{equation}\label{4ssB_ijmuzun}
B(i,j,m) =\displaystyle - \frac{A(i,j)P(p_m-p_i-p_j) +
A(i,m)P(p_j-p_i-p_m) + A(j,m)P(p_i-p_j-p_m)}{P(p_i+p_j+p_m)}
\end{equation}}}
\noindent for $i,j,m=1,2,3,4$ with $i<j<m$. From the coefficient
of $\varepsilon^4$ we have {{\small
\begin{equation}
\begin{split}
P(D)\{f_4.1+f_3.f_1+f_2.f_2+f_1.f_3+1.f_4
\}&=2P(\partial)f_4+2P(D)\{f_1.f_3 \}+P(D)\{f_2.f_2 \}\\
&=0.
\end{split}
\end{equation}}}
The simplifications gives us that we should have
\begin{equation}\label{4ssB_ijmkisa}
B(i,j,m)=A(i,j)A(i,m)A(j,m)
\end{equation}
for $i,j,m=1,2,3,4$ with $i<j<m$. To have consistency the
equations (\ref{4ssB_ijmuzun}) and (\ref{4ssB_ijmkisa}) should be
equivalent. This yields the condition
\begin{equation}
\sum_{\sigma_r=\pm
1}P(\sigma_ip_i+\sigma_jp_j+\sigma_mp_m)P(\sigma_ip_i-\sigma_jp_j)P(\sigma_jp_j-\sigma_mp_m)P(\sigma_ip_i-\sigma_mp_m)=0.
\end{equation}
for $i,j,m,r=1,2,3,4$ with $i<j<m$, which turns out to be
\begin{equation}\label{eKP3HCFOR4HS}
\begin{split}
&(12a-b^2)k_i^2k_j^2k_m^2\Big[2k_i^2w_jw_ml_jl_m+2k_j^2w_iw_ml_il_m+2k_m^2w_iw_jl_il_j
\\&+2k_jk_mw_jw_ml_i^2
+2k_ik_mw_iw_ml_j^2+2k_ik_jw_iw_jl_m^2+2k_jk_mw_i^2l_jl_m
\\&+2k_ik_mw_j^2l_il_m +2k_ik_jw_m^2l_il_j
-2k_ik_mw_jw_ml_il_j-2k_ik_mw_iw_jl_jl_m
\\&-2k_ik_jw_jw_ml_il_m-2k_ik_jw_iw_ml_jl_m
-2k_jk_mw_iw_ml_il_j-2k_jk_mw_iw_jl_il_m
\\&-k_i^2w_m^2l_j^2-k_i^2w_j^2l_m^2-k_j^2w_i^2l_m^2
-k_j^2w_m^2l_i^2-k_m^2w_j^2l_i^2-k_m^2w_i^2l_j^2\Big]=0
\end{split}
\end{equation}
where $i,j,m=1,2,3,4$, $i<j<m$. Some of the relations except
$a=b^2/12$ which make this condition satisfied are;\\

\noindent \textbf{Case 1.} Any two of $k_i=0$, $i=1,2,3,4$, the
rest are
different,\\

\noindent \textbf{Case 2.} $k_i=\omega_i$, $i=1,2,3,4$,\\

\noindent \textbf{Case 3.} $k_i=l_i$, $i=1,2,3,4$,\\

\noindent \textbf{Case 4.} $\omega_i=l_i$, $i=1,2,3,4$.\\

\noindent The equation remaining from the coefficient of
$\varepsilon^4$ is
\begin{multline}
-P(\partial)f_4=e^{\theta_1+\theta_2+\theta_3+\theta_4}[B_{123}P(p_4-p_1-p_2-p_3)+B_{124}P(p_3-p_1-p_2-p_4)]\\
+B_{134}P(p_2-p_1-p_3-p_4)+B_{234}P(p_1-p_2-p_3-p_4)+A(1,2)A(3,4)P(p_1+p_2-p_3-p_4)
\\+A(1,3)A(2,4)P(p_1+p_3-p_2-p_4)+A(1,4)A(2,3)P(p_1+p_4-p_2-p_3)]=0.
\end{multline}
Thus $f_4=Ce^{\theta_1+\theta_2+\theta_3+\theta_4}$ where $C$ is
obtained as
\begin{equation}\label{eKPCuzun}
\begin{split}
C=-[&A(1,2)A(3,4)P(p_1+p_2-p_3-p_4)
+A(1,3)A(2,4)P(p_1+p_3-p_2-p_4)\\+&A(1,4)A(2,3)P(p_1+p_4-p_2-p_3)
+B(1,2,3)P(p_4-p_1-p_2-p_3)\\+&B(1,2,4)P(p_3-p_1-p_2-p_4)
+B(1,3,4)P(p_2-p_1-p_3-p_4)\\+&B(2,3,4)P(p_1-p_2-p_3-p_4) ]\Bigg
/P(p_1+p_2+p_3+p_4).
\end{split}
\end{equation}
\noindent By the coefficient of $\varepsilon^5$ we have
\begin{equation}
2P(\partial)f_4+2P(D)\{f_1.f_3 \}+P(D)\{f_2.f_2 \}=0
\end{equation}
and when we put the expressions that we have found for $f_i$,
$i=1,2,3,4$ into this equation, it gives
\begin{equation}\label{eKPCkisa}
C=A(1,2)A(1,3)A(1,4)A(2,3)A(2,4)A(3,4).
\end{equation}
To be consistent the equations (\ref{eKPCuzun}) and
(\ref{eKPCkisa}) should be equal to each other. This yields the
condition
\begin{equation}
\begin{split}
&P(p_1-p_2)P(p_1-p_3)P(p_2-p_3)P(p_1+p_4)\\
&\hspace{43mm} \times P(p_2+p_4)P(p_3+p_4)P(p_4-p_1-p_2-p_3)\\
&+P(p_1-p_2)P(p_1-p_4)P(p_2-p_4)P(p_1+p_3)\\
&\hspace{43mm}\times P(p_2+p_3)P(p_3+p_4)P(p_3-p_1-p_2-p_4)\\
&+P(p_1-p_3)P(p_1-p_4)P(p_3-p_4)P(p_1+p_2)
\\ &\hspace{43mm}\times P(p_2+p_3)P(p_2+p_4)P(p_2-p_1-p_3-p_4)\\
&+P(p_2-p_3)P(p_2-p_4)P(p_3-p_4)P(p_1+p_2)\\
&\hspace{43mm}\times P(p_1+p_3)P(p_1+p_4)P(p_1-p_2-p_3-p_4)\\
&-P(p_1-p_2)P(p_3-p_4)P(p_1+p_3)P(p_1+p_4)
\\ &\hspace{43mm}\times P(p_2+p_3)P(p_2+p_4)P(p_1+p_2-p_3-p_4)\\
&-P(p_1-p_3)P(p_2-p_4)P(p_1+p_2)P(p_1+p_4)
\\ &\hspace{43mm}\times P(p_2+p_3)P(p_3+p_4)P(p_1+p_3-p_2-p_4)\\
&-P(p_1-p_4)P(p_2-p_3)P(p_1+p_2)P(p_1+p_3)\\
&\hspace{43mm}\times P(p_2+p_4)P(p_3+p_4)P(p_1+p_4-p_2-p_3)\\
&-P(p_1-p_2)P(p_1-p_3)P(p_1-p_4)P(p_2-p_3)\\
&\hspace{43mm}\times P(p_2-p_4)P(p_3-p_4)P(p_1+p_2+p_3+p_4)=0.
\end{split}
\end{equation}
which can also be written as
\begin{equation}
\sum_{\sigma_i=\pm 1}P(\sum_{i=1}^4 \sigma_i p_i) \prod_{0<i<j<4}
[P(\sigma_i p_i-\sigma_j p_j)] = 0.
\end{equation}
\noindent We call this condition as four-Hirota solution condition
$(4HC)$. Here the question is whether the cases satisfying
(\ref{eKP3HCFOR4HS}) also satisfy $(4HC)$ automatically. In the
hand we know a case which satisfies both conditions which is\\

\noindent \textbf{Case 1.} Any two of $k_i=0$, $i=1,2,3,4$, the
rest are different.\\

\noindent Now, for illustration, let us see the graphs of the two-
and four-Hirota solutions of eKP. Here in order to get the
solutions we give arbitrary values to $a$, $b$, $k_i$ and $w_i$
and from the dispersion relation we obtain $l_i$. Note that in our
choice $a\neq b^2/12$.\\

\noindent \textbf{i)} \textbf{$N=2$, The Two-Hirota Solution of EKP}:\\

\noindent The constants are\\

\noindent $a=2$, $b=9$, $k_1=1$, $k_2=3$,\\

\noindent $w_1=4$, $w_2=-5$, $\displaystyle
l_1=-6+\frac{\sqrt{213}}{3}$, $\displaystyle
l_2=\frac{15}{2}+\frac{\sqrt{633}}{2}$.\\

\begin{center}
\begin{figure}[h]
\centering
\begin{minipage}[t]{0.4\linewidth}
\setlength{\fboxsep}{-\fboxrule}

\fbox{\includegraphics[angle=270,scale=.28]{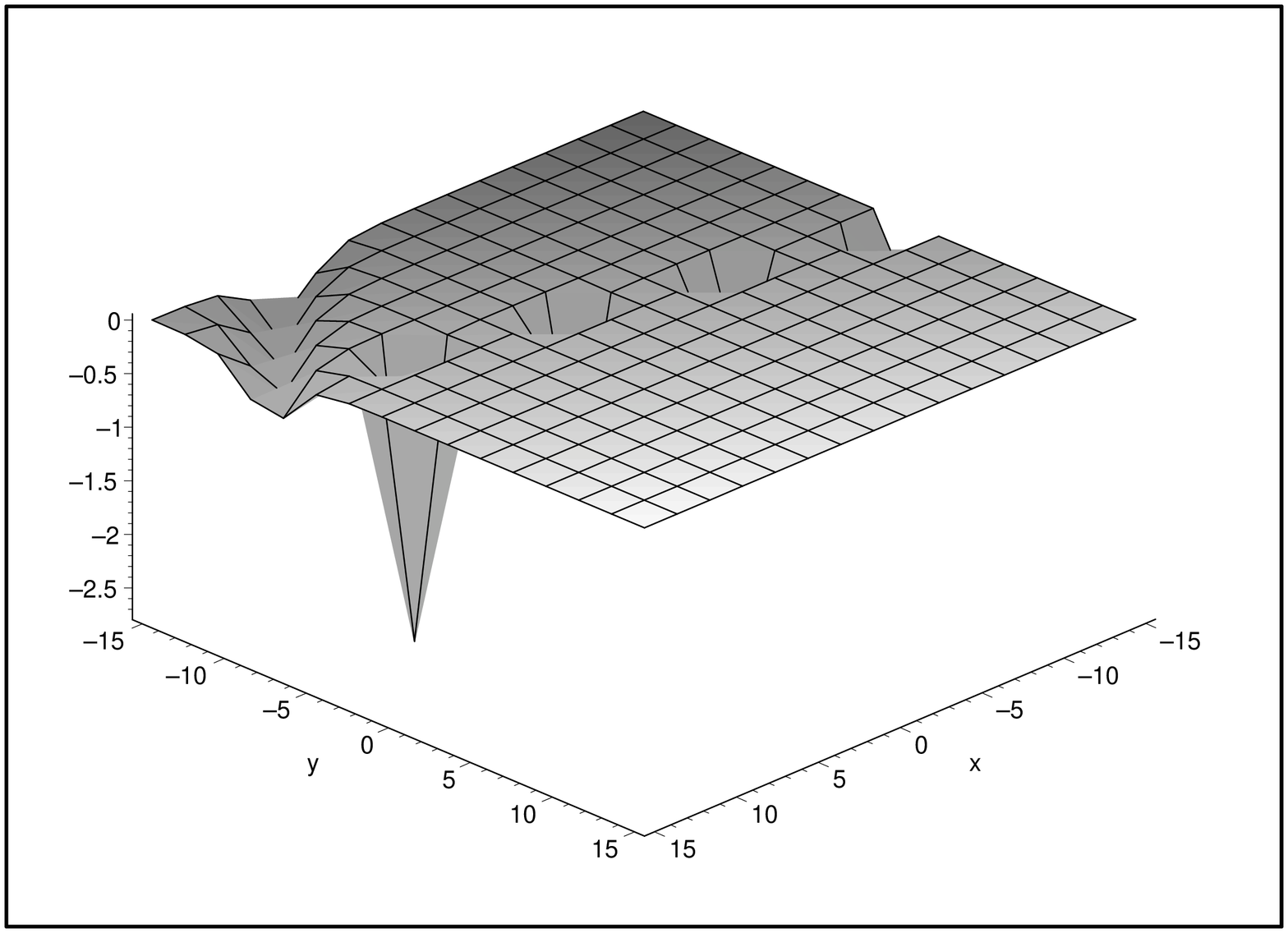}}
\caption{t=-6}
\end{minipage}%
\hspace{1cm}%
\begin{minipage}[t]{0.4\linewidth}
\setlength{\fboxsep}{-\fboxrule}
\fbox{\includegraphics[angle=270,scale=.28]{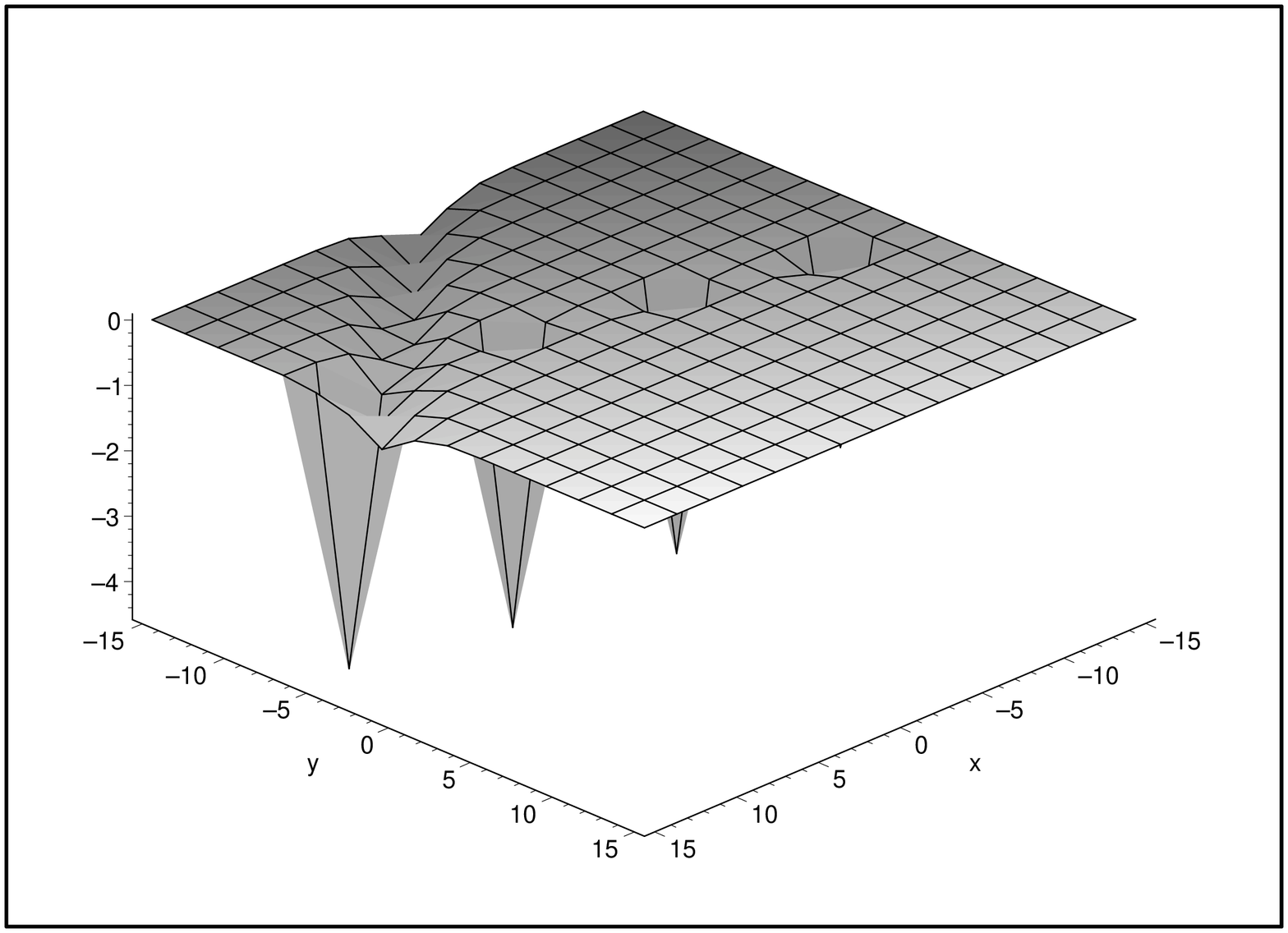}}
\caption{t=-4}
\end{minipage}
\end{figure}
\begin{figure}[h]
\centering
\begin{minipage}[t]{0.4\linewidth}
\setlength{\fboxsep}{-\fboxrule}
\fbox{\includegraphics[angle=270,scale=.28]{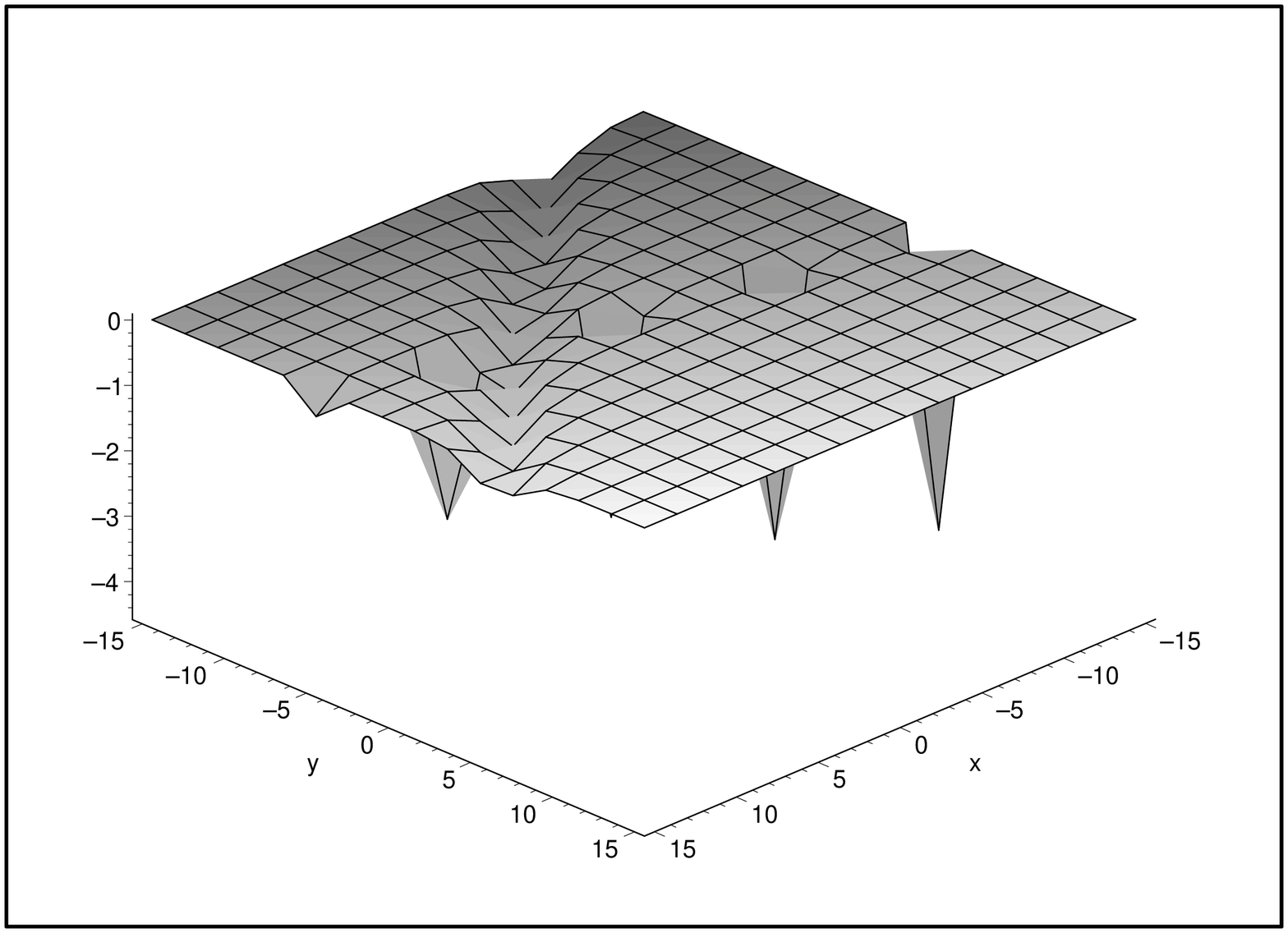}}
\caption{t=-2}
\end{minipage}%
\hspace{1cm}%
\begin{minipage}[t]{0.4\linewidth}
\setlength{\fboxsep}{-\fboxrule}
\fbox{\includegraphics[angle=270,scale=.28]{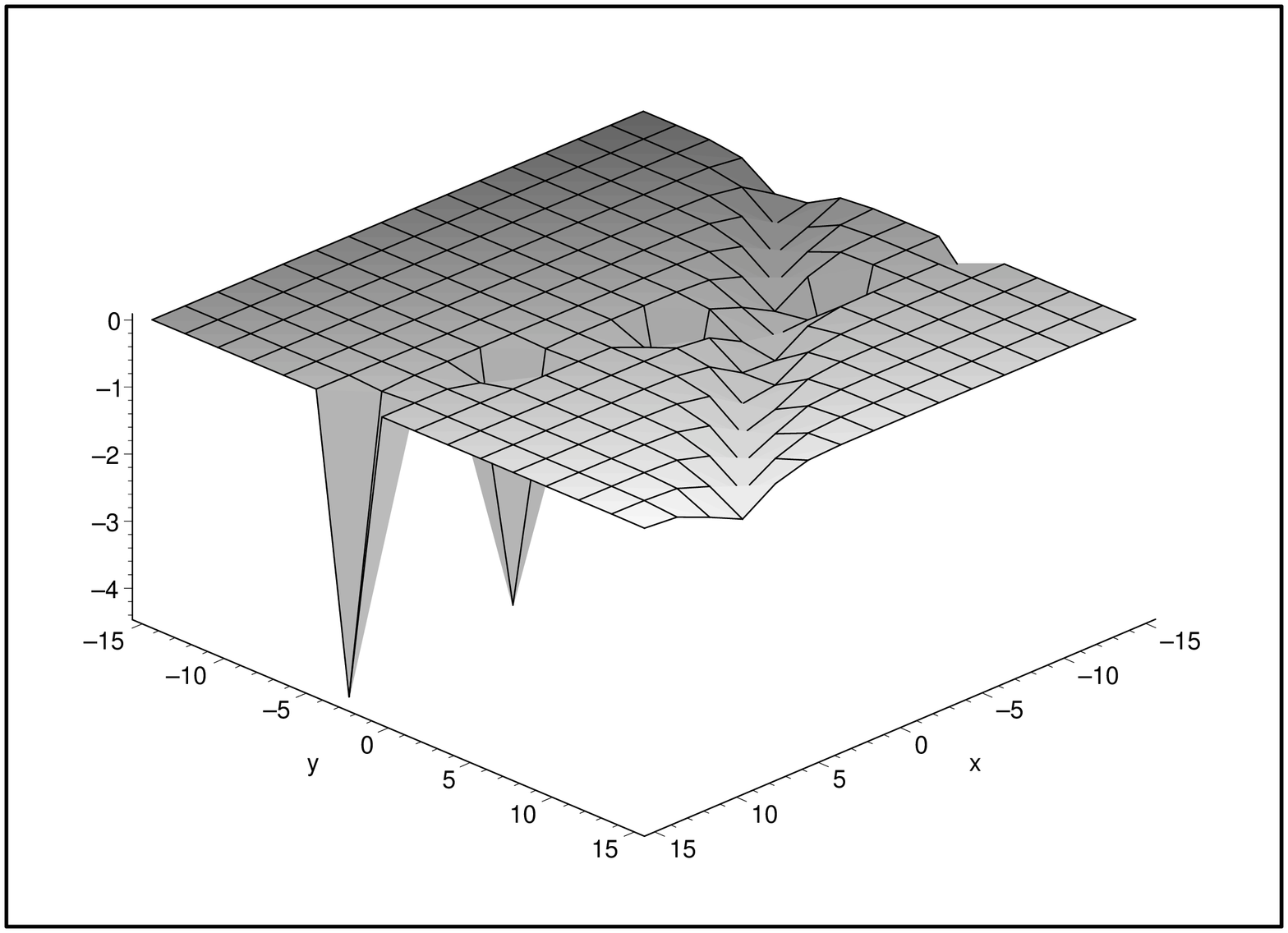}}
\caption{t=2}
\end{minipage}
\end{figure}

\newpage

\begin{figure}[h]
\centering
\begin{minipage}[t]{0.4\linewidth}
\setlength{\fboxsep}{-\fboxrule}
\fbox{\includegraphics[angle=270,scale=.28]{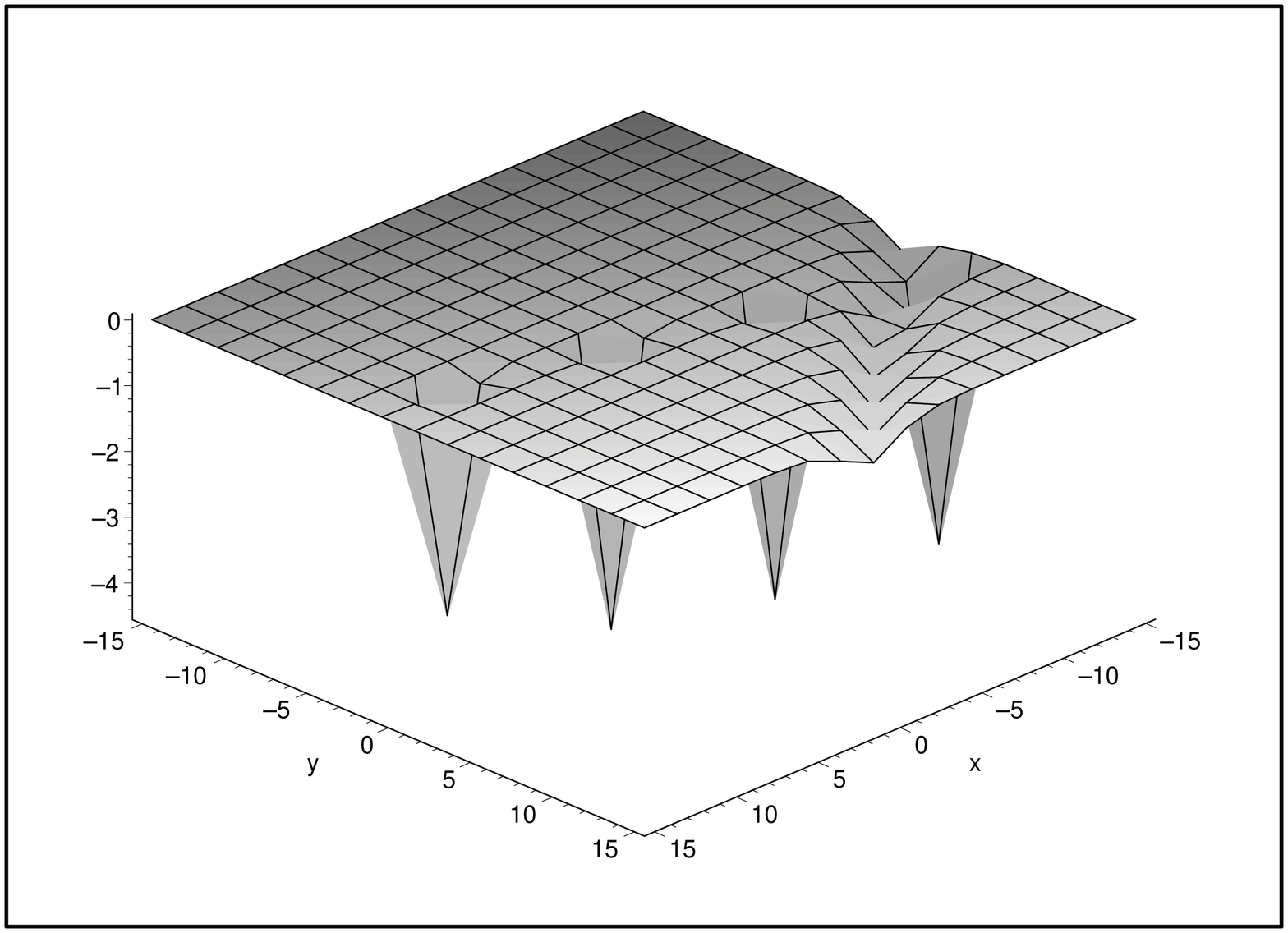}}
\caption{t=4}
\end{minipage}%
\hspace{1cm}%
\begin{minipage}[t]{0.4\linewidth}
\setlength{\fboxsep}{-\fboxrule}
\fbox{\includegraphics[angle=270,scale=.28]{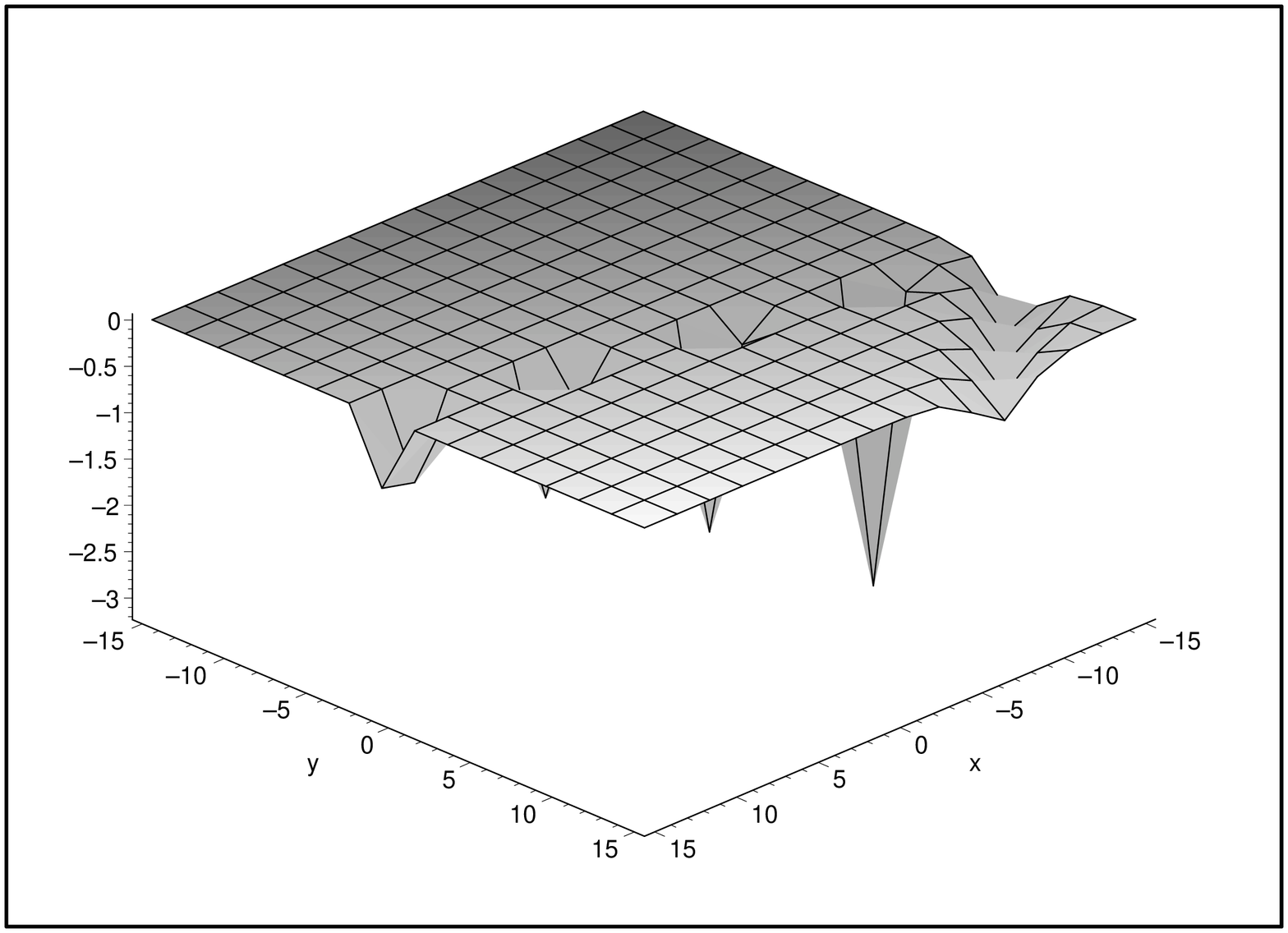}}
\caption{t=6}
\end{minipage}
\end{figure}
\end{center}

\noindent \textbf{ii)} \textbf{$N=4$, The Four-Hirota Solution of EKP}:\\

\noindent The constants are chosen according to the \textbf{Case
1} and the dispersion relation. The constants are,\\

\noindent $a=2$, $b=9$, $k_1=0$, $k_2=0$, $k_3=1$, $k_4=2$,\\

\noindent $w_1=4$, $w_2=-2$, $w_3=3$, $w_4=-5$,\\

\noindent $\displaystyle l_1=-6+\frac{2\sqrt{57}}{3}$,
$\displaystyle l_2=3+\frac{\sqrt{57}}{3}$, $\displaystyle
l_3=-\frac{9}{2}+\frac{\sqrt{465}}{6}$, $\displaystyle
l_4=\frac{15}{2}+\frac{\sqrt{1353}}{6}$.

\begin{center}
\begin{figure}[h]
\centering
\begin{minipage}[t]{0.4\linewidth}
\setlength{\fboxsep}{-\fboxrule}
\fbox{\includegraphics[angle=270,scale=.28]{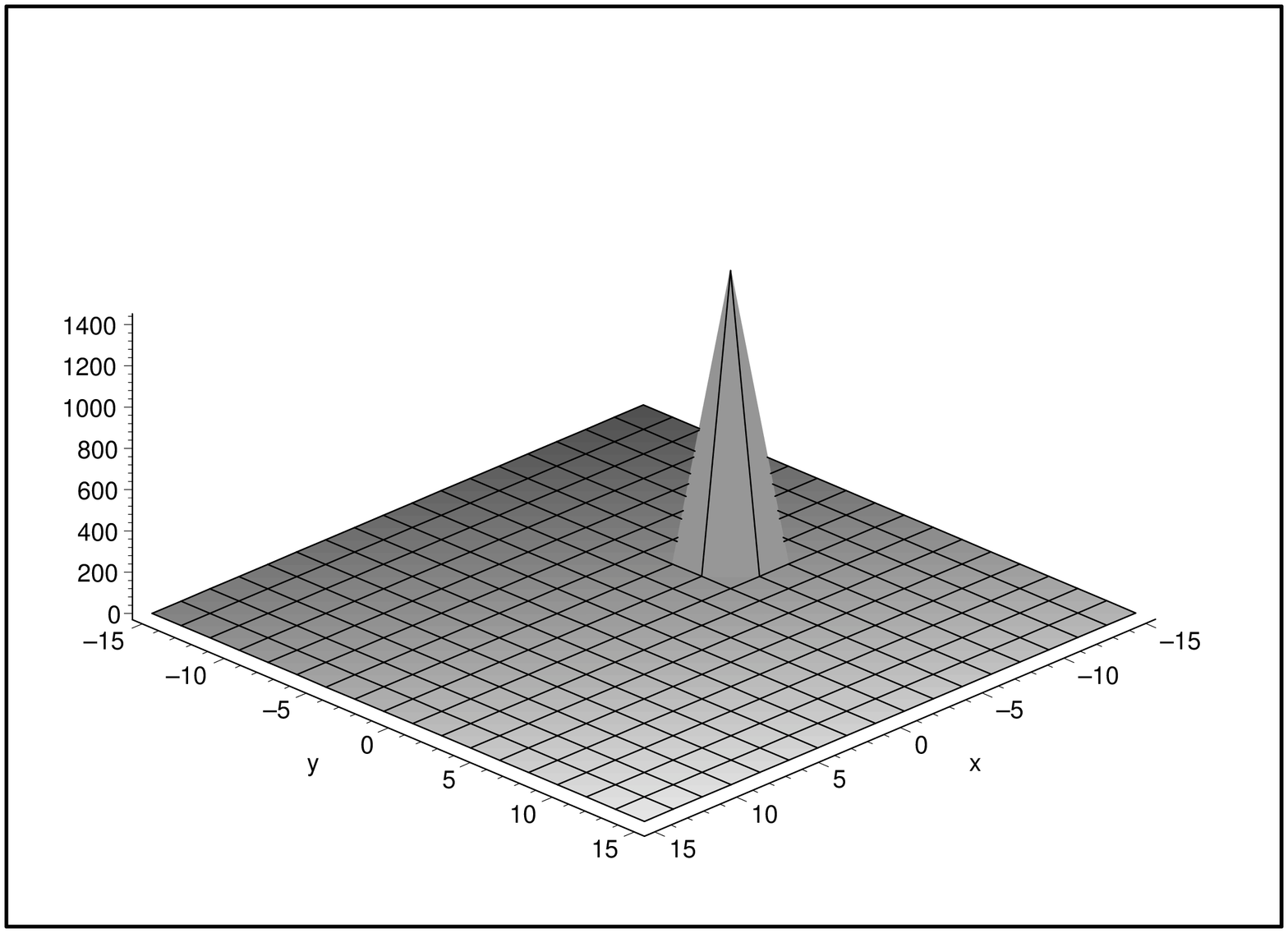}}
\caption{t=-6}
\end{minipage}%
\hspace{1cm}%
\begin{minipage}[t]{0.4\linewidth}
\setlength{\fboxsep}{-\fboxrule}
\fbox{\includegraphics[angle=270,scale=.28]{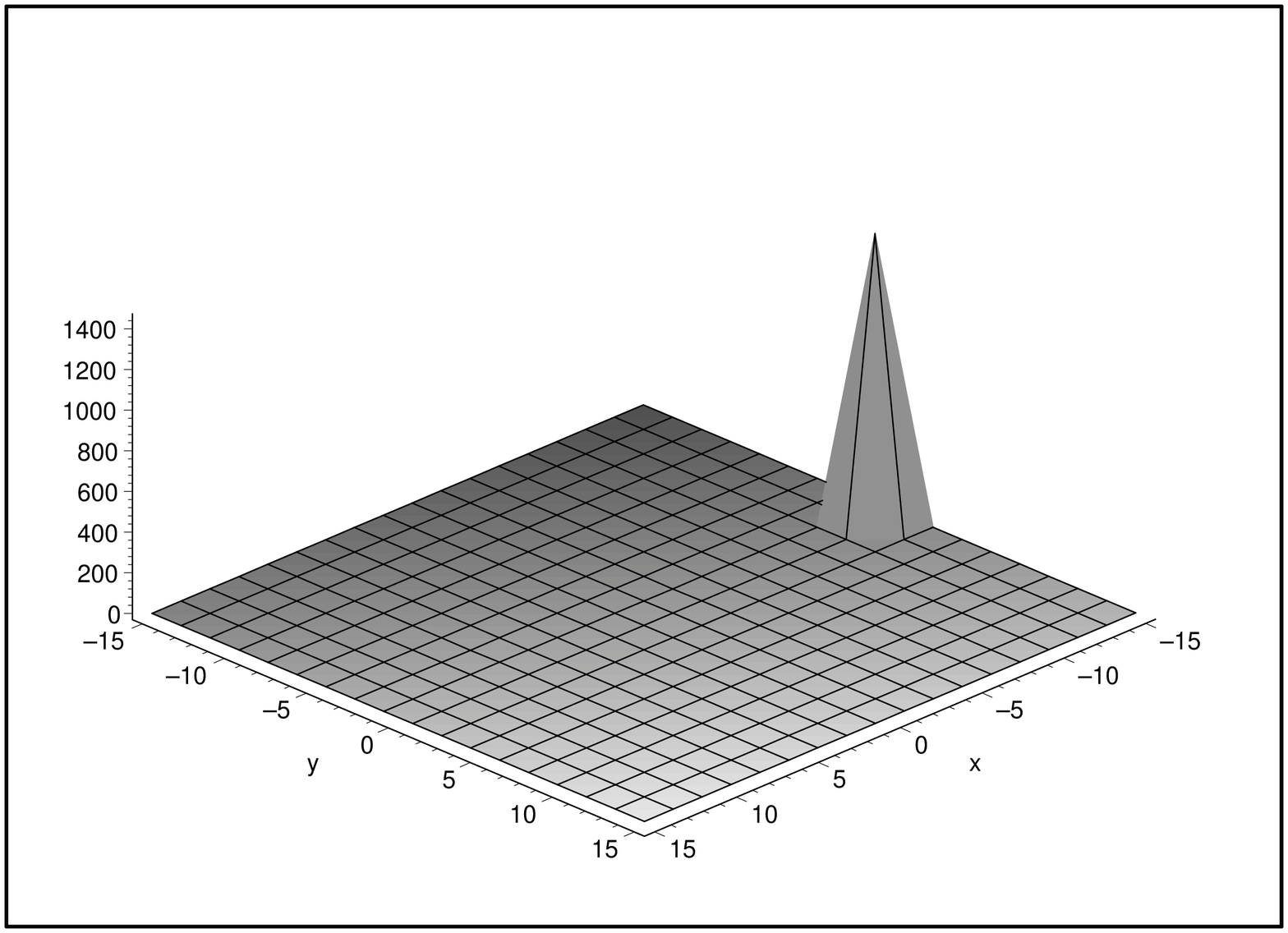}}
\caption{t=-4}
\end{minipage}
\end{figure}

\newpage
\begin{figure}[h]
\centering
\begin{minipage}[t]{0.4\linewidth}
\setlength{\fboxsep}{-\fboxrule}
\fbox{\includegraphics[angle=270,scale=.28]{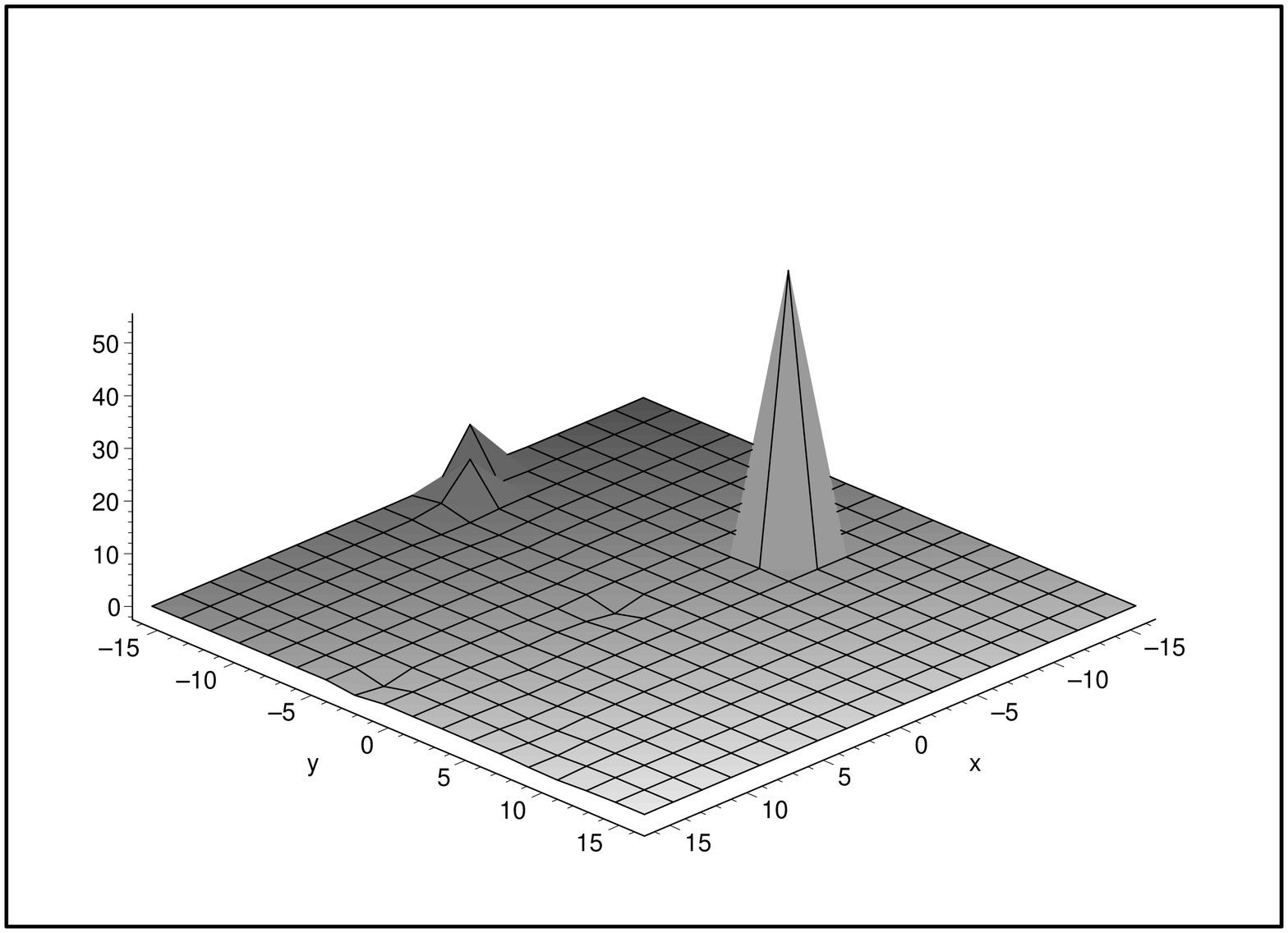}}
\caption{t=-2}
\end{minipage}%
\hspace{1cm}%
\begin{minipage}[t]{0.4\linewidth}
\setlength{\fboxsep}{-\fboxrule}
\fbox{\includegraphics[angle=270,scale=.28]{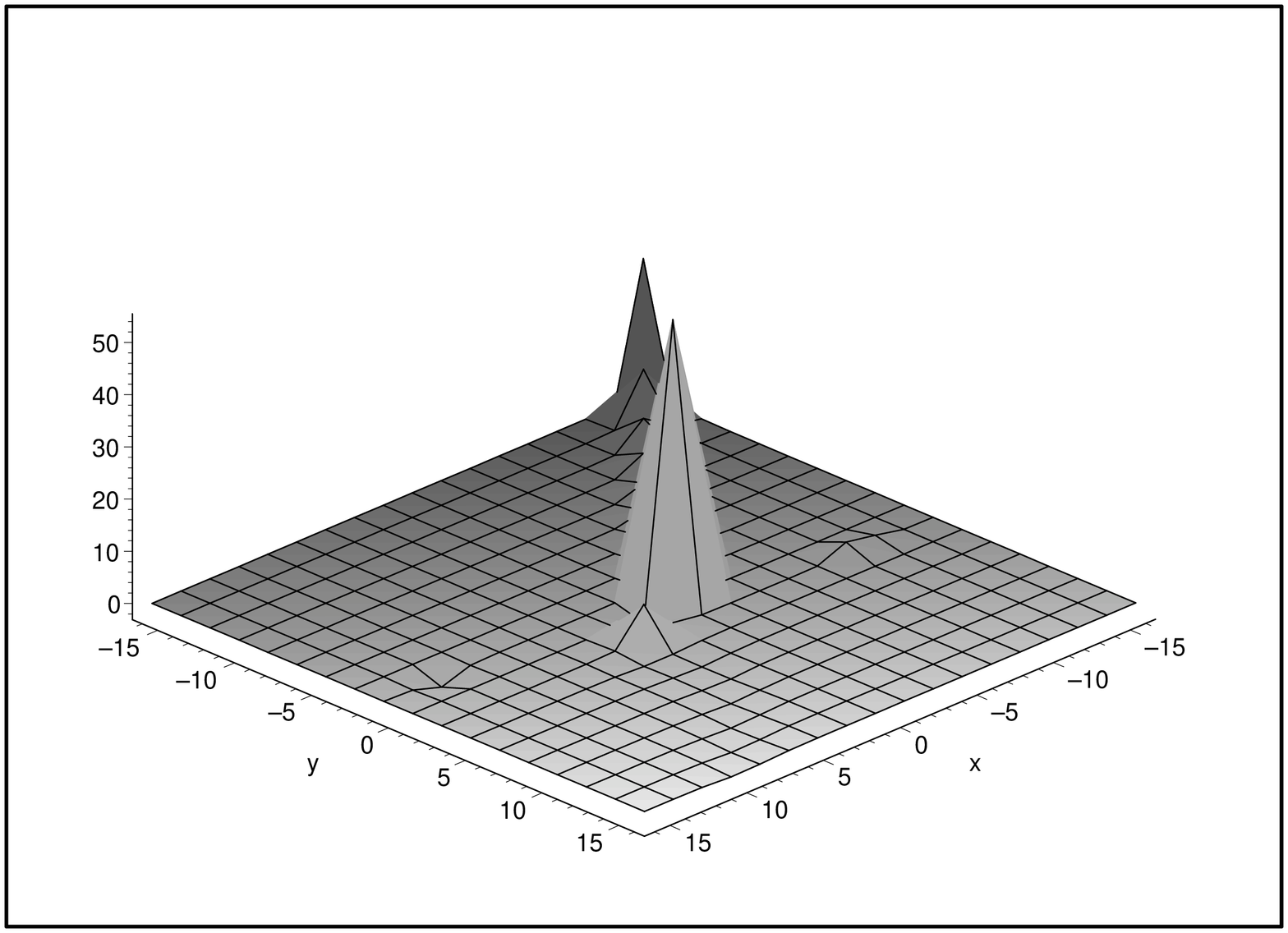}}
\caption{t=2}
\end{minipage}
\end{figure}
\begin{figure}[h]
\centering
\begin{minipage}[t]{0.4\linewidth}
\setlength{\fboxsep}{-\fboxrule}
\fbox{\includegraphics[angle=270,scale=.28]{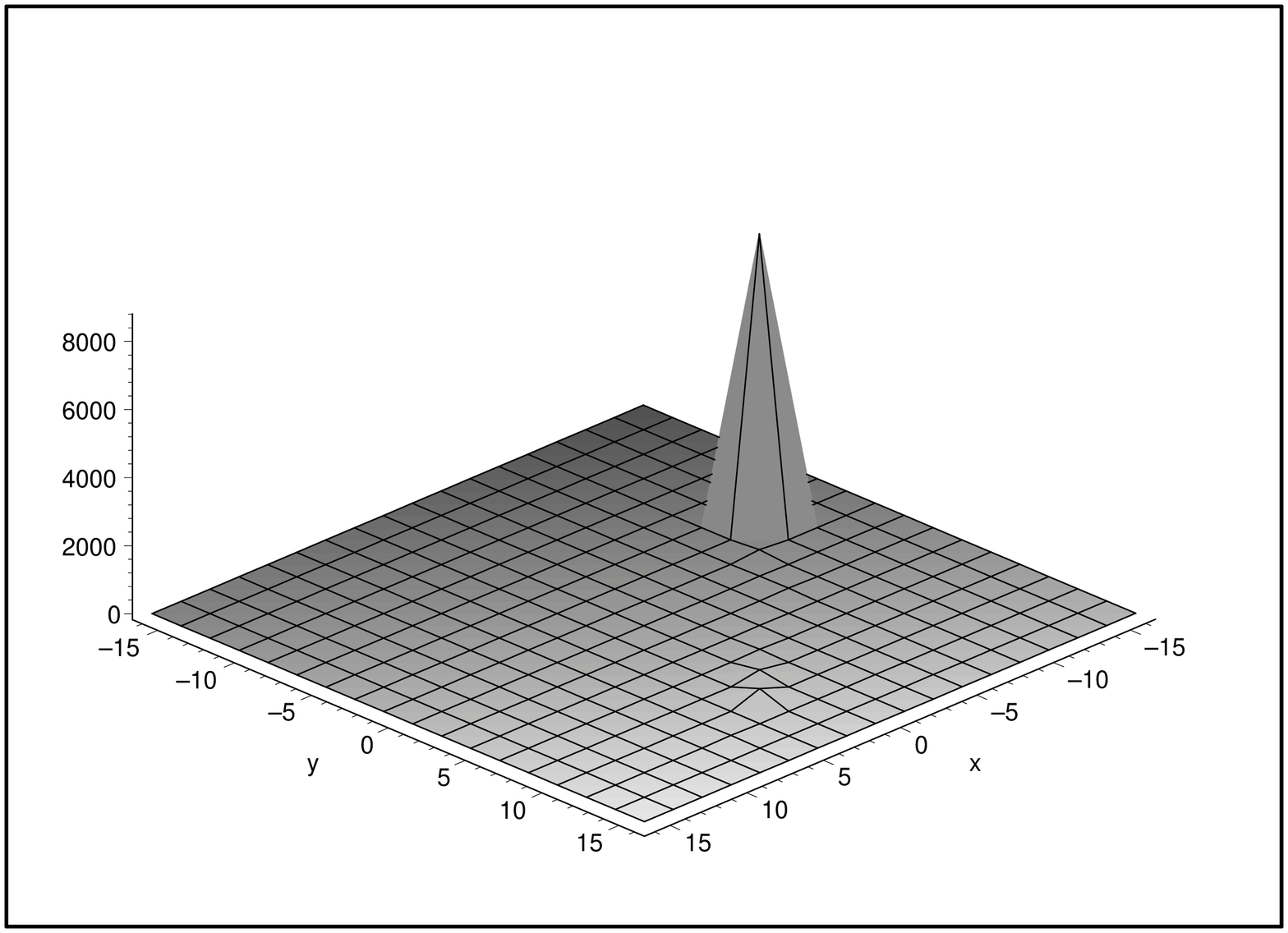}}
\caption{t=4}
\end{minipage}%
\hspace{1cm}%
\begin{minipage}[t]{0.4\linewidth}
\setlength{\fboxsep}{-\fboxrule}
\fbox{\includegraphics[angle=270,scale=.28]{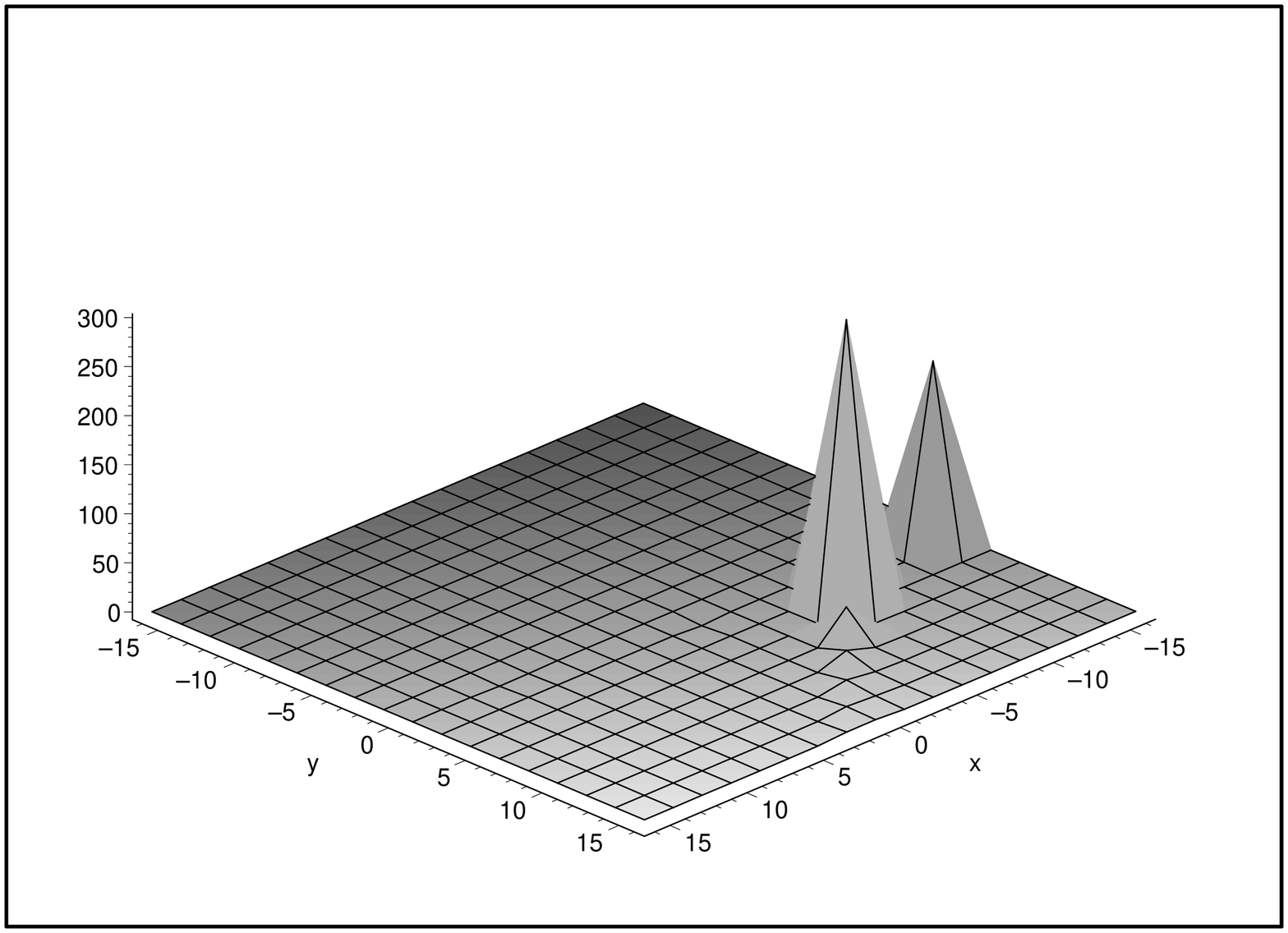}}
\caption{t=6}
\end{minipage}
\end{figure}
\end{center}

\vspace{4mm}

\subsection{The Extended Boussinesq (EBo) Equation}

The extended Boussinesq (eBo) equation is given by
\begin{equation}
u_{tt}-u_{xx}+3(u^2)_{xx}-u_{xxxx}+au_{yy}+bu_{ty}+c\nabla^2 u=0
\end{equation}
which is constructed by adding the terms $au_{yy}$, $bu_{ty}$ and
$\nabla^2 u=u_{x_1x_1}+u_{x_2x_2}+...+u_{x_mx_m}$ multiplied by
$c$ to the Boussinesq (Bo) equation where $a$, $b$ and $c$ are
constants and $x_j$, $j=1,2,...,m$ are independent variables. Now
let us apply the Hirota direct method to the eBo equation.
\noindent \textbf{Step $1$}. \textit{Bilinearization}: We use the
bilinearizing transformation
\begin{equation}
u(x,t,y)=-2\partial_x^2 \log f
\end{equation}
\noindent so the bilinear form of eBo is
\begin{multline}\label{eBbilinear}
ff_{tt}-f_t^2-f_{xx}f+f_x^2-f_{xxxx}f+4f_xf_{xxx}-3f_{xx}^2\\
+af_{yy}f-af_y^2+bf_{ty}f-bf_tf_y+c\sum_{j=1}^m(f_{x_jx_j}f-f_{x_j}^2)
=0.
\end{multline}
\noindent \textbf{Step $2$}. \textit{Transformation to the Hirota
bilinear form}: The Hirota bilinear form of eBo is
\begin{equation}\label{eBhirotabilinear}
P(D)\{f.f
\}=(D_t^2-D_x^2-D_x^4+aD_y^2+bD_tD_y+c\sum_{j=1}^mD_{x_j}^2)\{f.f
\}=0.
\end{equation}

\noindent \textbf{Step $3$.} \textit{Application of the Hirota
perturbation}: Insert $f=1+\sum_{n=1}^N \varepsilon^n f_n$into the
equation (\ref{eBhirotabilinear}) so we have
\begin{multline}\label{EBPERTURBATION}
P(D)\{f.f \}=P(D)\{1.1 \} + \varepsilon P(D)\{f_1.1 +1.f_1
\}\\+...+\varepsilon^{2N} P(D)\{f_N.f_N \}=0.
\end{multline}

\noindent \textbf{Step $4$}: \textit{Examination of the
coefficients of the perturbation parameter $\varepsilon$}: We make
the coefficients of $\varepsilon^m$, $m=1,2,...,N$ appeared in
(\ref{EBPERTURBATION}) to vanish. Here we shall consider only the
case $N=3$ and $N=4$.

\subsubsection{$N=3$, Three-Hirota Solution of EBo}
 \noindent Here we apply the Hirota direct method by using the
 ansatz which is used to construct three-soliton solutions.
 We take $ f = 1 + \varepsilon f_1 + \varepsilon^2 f_2 +
\varepsilon^3 f_3$ where $f_1 = e^{\theta_1}+e^{\theta_2} +
e^{\theta_3}$ with $\theta_i = k_ix + \omega_it + l_iy +
\sum_{j=1}^m r_{ij}x_j+\alpha_i$ for $i = 1,2,3$ and insert it
into (\ref{EBPERTURBATION}). The coefficient of $\varepsilon^0$ is
identically zero. By the coefficient of $\varepsilon^1$
\begin{equation}
P(D)\{1.f_1 + f_1.1 \}
=2P(\partial)\{e^{\theta_1}+e^{\theta_2}+e^{\theta_3}\}=0
\end{equation} we have the dispersion relation
\begin{equation}\label{eBdispersion}
P(p_i)=\omega_i^2-k_i^2-k_i^4+al_i^2+b \omega_il_i+c\sum_j^m
r_{ij}^2=0
\end{equation}
for $i=1,2,3$. Similar to the eKP equation, we see that when $c$,
the coefficient of $\nabla^2 u$ is not zero, we can apply the
Hirota method. But for simplicity we take $c=0$ in the rest of the
calculations. In this case $\theta_i$ become $\theta_i = k_ix +
\omega_it + l_iy + \alpha_i$. From the coefficient of
$\varepsilon^2$ we get
\begin{multline}\label{eB3ssf_2}
-P(\partial)f_2=\sum_{i<j}^{(3)}[(\omega_i-\omega_j)^2-(k_i-k_j)^2-(k_i-k_j)^4+a(l_i-l_j)^2\\
+b(\omega_i-\omega_j)(l_i-l_j)]e^{\theta_i+\theta_j}
\end{multline}
\noindent where $(3)$ indicates the summation of all possible
combinations of the three elements with
 $i<j$. Thus $f_2$ should be of the form
 \begin{equation}
f_2 = A(1,2)e^{{\theta_1}+{\theta_2}} +
A(1,3)e^{{\theta_1}+{\theta_3}} + A(2,3)e^{{\theta_2}+{\theta_3}}
\end{equation}
to satisfy the equation. We insert $f_2$ into the equation
(\ref{eB3ssf_2}) so we get $A(i,j)$ as
\begin{equation}
\begin{split}
 A(i,j)&=-\frac{P(p_i-p_j)}{P(p_i+P_j)}\\
 &=\displaystyle
 \frac{2\omega_i\omega_j-2k_ik_j-4k_i^3k_j+6k_i^2k_j^2-4k_ik_j^3+2al_il_j+b\omega_il_j
 +b\omega_jl_i
 }{2\omega_i\omega_j-2k_ik_j-4k_i^3k_j-6k_i^2k_j^2-4k_ik_j^3+2al_il_j+b\omega_il_j+b\omega_jl_i}
\end{split}
\end{equation}
\noindent where $i,j=1,2,3$, $i<j$. From the coefficient of
$\varepsilon^3$, we obtain
\begin{multline}
P(\partial)\{f_3 \} = -[A(1,2)P(p_3-p_2-p_1)+A(1,3)P(p_2-p_1-p_3)\\
+A(2,3)P(p_1-p_2-p_3)]e^{{\theta_1}+{\theta_2} + {\theta_3}}.
\end{multline}
Hence $f_3$ is in the form $f_3 =Be^{{\theta_1}+{\theta_2} +
{\theta_3}}$ where $B$ is found as {{\small
\begin{equation}\label{eBBuzun}
B =\displaystyle - \frac{A(1,2)P(p_3-p_1-p_2) +
A(1,3)P(p_2-p_1-p_3) + A(2,3)P(p_1-p_2-p_3)}{P(p_1+p_2+p_3)}.
\end{equation}}}
\noindent The coefficient of $\varepsilon^4$ gives us
\begin{multline}
e^{{2\theta_1}+{\theta_2}+{\theta_3}}
[BP(p_2+p_3)+A(1,2)A(1,3)P(p_2-p_3)]\\+
e^{{\theta_1}+{2\theta_2}+{\theta_3}}
[BP(p_1+p_3)+A(1,2)A(2,3)P(p_1-p_3)]\\+
e^{{\theta_1}+{\theta_2}+{2\theta_3}}
[BP(p_1+p_2)+A(1,3)A(2,3)P(p_1-p_2)]=0
\end{multline}
which is satisfied when
\begin{equation}\label{eBBkisa}
B = A(1,2)A(1,3)A(2,3).
\end{equation}
The two expressions for $B$ should be equivalent. This yields the
three-Hirota solution condition $(3HC)$
\begin{equation}
\sum_{\sigma_r=\pm
1}P(\sigma_1p_1+\sigma_2p_2+\sigma_3p_3)P(\sigma_1p_1-\sigma_2p_2)P(\sigma_2p_2-\sigma_3p_3)P(\sigma_1p_1-\sigma_3p_3)=0,
\end{equation}
$r=1,2,3$ which turns out to be the below equation for eBo,
\begin{equation}
\begin{split}
&(4a-b^2)k_1^2k_2^2k_3^2\Big[2k_1^2w_2w_3l_2l_3+2k_2^2w_1w_3l_1l_3+2k_3^2w_1w_2l_1l_2
\\&+2k_2k_3w_2w_3l_1^2
+2k_1k_3w_1w_3l_2^2+2k_1k_2w_1w_2l_3^2+2k_2k_3w_1^2l_2l_3
\\&+2k_1k_3w_2^2l_1l_3 +2k_1k_2w_3^2l_1l_2
-2k_1k_3w_2w_3l_1l_2-2k_1k_3w_1w_2l_2l_3
\\&-2k_1k_2w_2w_3l_1l_3-2k_1k_2w_1w_3l_2l_3
-2k_2k_3w_1w_3l_1l_2-2k_2k_3w_1w_2l_1l_3
\\&-k_1^2w_3^2l_2^2-k_1^2w_2^2l_3^2-k_2^2w_1^2l_3^2
-k_2^2w_3^2l_1^2-k_3^2w_2^2l_1^2-k_3^2w_1^2l_2^2\Big]=0.
\end{split}
\end{equation}
It is satisfied when $a=b^2/4$ or for some relations on $k_i$,
$w_i$ and $l_i$ which violate the solitonic property of the
solution. The coefficients of $\varepsilon^5$ and $\varepsilon^6$
vanish trivially. Similar to eKP, when $a=b^2/4$, eBo is
integrable since it is transformable to Bo by the transformation
\begin{center}
$u'=u$, $x'=x$, $y'=\alpha t+\beta y$, $t'=t+ \rho y$,
\end{center}
\noindent where $\alpha=b/2=\sqrt{a}$. When $a\neq b^2/4$, eBo is
not integrable. The other relations which makes $(3HC)$ satisfied
are;
\\

\noindent \textbf{Case 1.} Any $k_i=0$, $i=1,2,3$, the rest are
different,\\

\noindent \textbf{Case 2.} $k_i=\omega_i$, $i=1,2,3$,\\

\noindent \textbf{Case 3.} $k_i=l_i$, $i=1,2,3$,\\

\noindent \textbf{Case 4.} $\omega_i=l_i$, $i=1,2,3$.\\

\noindent By using any of these cases, we obtain the exact
solutions of eBo.

\subsubsection{$N=4$, Four-Hirota Solution of EBo}
 \noindent Here we apply the Hirota direct method by using the
 ansatz which is used to construct four-soliton solutions. We
 take $ f = 1 + \varepsilon f_1 + \varepsilon^2 f_2 +
\varepsilon^3 f_3+\varepsilon^4 f_4$ where $f_1 =
e^{\theta_1}+e^{\theta_2} + e^{\theta_3}+e^{\theta_4}$ with
$\theta_i = k_ix + \omega_it + l_iy + \alpha_i$ for $i = 1,2,3,4$
and insert it into (\ref{EBPERTURBATION}). We will only consider
the coefficients of $\varepsilon^m$, $m=1, 2, 3, 4, 5$ since the
others vanish identically. By the coefficient of $\varepsilon^1$
\begin{equation}
P(D)\{1.f_1 + f_1.1 \}
=2P(\partial)\{e^{\theta_1}+e^{\theta_2}+e^{\theta_3}+e^{\theta_4}
\}=0
\end{equation} we have the dispersion relation
\begin{equation}
P(p_i)=\omega_i^2-k_i^2-k_i^4+al_i^2+b \omega_il_i=0
\end{equation}
for $i=1,2,3,4$. From the coefficient of $\varepsilon^2$ we obtain
\begin{multline}
-P(\partial)f_2=P(p_1-p_2)e^{\theta_1+\theta_2}+P(p_1-p_3)e^{\theta_1+\theta_3}+P(p_1-p_4)e^{\theta_1+\theta_4}\\
+P(p_2-p_3)e^{\theta_2+\theta_3}+P(p_2-p_4)e^{\theta_2+\theta_4}+P(p_3-p_4)e^{\theta_3+\theta_4}.
\end{multline}
\noindent Thus $f_2$ should be of the form
\begin{multline}
f_2 = A(1,2)e^{{\theta_1}+{\theta_2}} +
A(1,3)e^{{\theta_1}+{\theta_3}} +A(1,4)e^{\theta_1+\theta_4}\\
+
A(2,3)e^{{\theta_2}+{\theta_3}}+A(2,4)e^{\theta_2+\theta_4}+A(3,4)e^{\theta_3+\theta_4},
\end{multline}
where $A(i,j)$ are obtained as
\begin{equation}
\begin{split}
 A(i,j)&=-\frac{P(p_i-p_j)}{P(p_i+P_j)}\\
 &=\displaystyle
 \frac{2\omega_i\omega_j-2k_ik_j-4k_i^3k_j+6k_i^2k_j^2-4k_ik_j^3+2al_il_j+b\omega_il_j
 +b\omega_jl_i
 }{2\omega_i\omega_j-2k_ik_j-4k_i^3k_j-6k_i^2k_j^2-4k_ik_j^3+2al_il_j+b\omega_il_j+b\omega_jl_i}
\end{split}
\end{equation}
\noindent for $i,j=1,2,3,4$ with $i<j$. The coefficient of
$\varepsilon^3$ gives
\begin{multline}\label{EB4ssf_3}
-P(\partial)f_3
=\sum_{i<j<m}^{(4)}[A(i,j)P(p_m-p_i-p_j)+A(i,m)P(p_j-p_i-p_m)\\
+A(j,m)P(p_i-p_j-p_m)]e^{\theta_i+\theta_j+\theta_m}
\end{multline}
\noindent where $(4)$ indicates the summation of all possible
combinations of the four elements with
 $i<j<m$. Hence $f_3$ is of the form
\begin{multline}
f_3 =B(1,2,3)e^{{\theta_1}+{\theta_2} +
{\theta_3}}+B(1,2,4)e^{{\theta_1}+{\theta_2} +
{\theta_4}}\\+B(1,3,4)e^{{\theta_1}+{\theta_3} +
{\theta_4}}+B(2,3,4)e^{{\theta_2}+{\theta_3} + {\theta_4}}.
\end{multline}
$B(i,j,m)$ are obtained as {{\small
\begin{equation}\label{EB4ssB_ijmuzun}
B(i,j,m) =\displaystyle - \frac{A(i,j)P(p_m-p_i-p_j) +
A(i,m)P(p_j-p_i-p_m) + A(j,m)P(p_i-p_j-p_m)}{P(p_i+p_j+p_m)}
\end{equation}}}
\noindent for $i,j,m=1,2,3,4$ with $i<j<m$. From the coefficient
of $\varepsilon^4$ we have {{\small
\begin{equation}
\begin{split}
P(D)\{f_4.1+f_3.f_1+f_2.f_2+f_1.f_3+1.f_4
\}&=2P(\partial)f_4+2P(D)\{f_1.f_3 \}+P(D)\{f_2.f_2 \}\\
&=0.
\end{split}
\end{equation}}}
After simplifications we see that we should have
\begin{equation}\label{EB4ssB_ijmkisa}
B(i,j,m)=A(i,j)A(i,m)A(j,m)
\end{equation}
for $i,j,m=1,2,3,4$ with $i<j<m$. To be consistent the equations
(\ref{EB4ssB_ijmuzun}) and (\ref{EB4ssB_ijmkisa}) should be
equivalent. This gives us the condition
\begin{equation}
\sum_{\sigma_r=\pm
1}P(\sigma_ip_i+\sigma_jp_j+\sigma_mp_m)P(\sigma_ip_i-\sigma_jp_j)P(\sigma_jp_j-\sigma_mp_m)P(\sigma_ip_i-\sigma_mp_m)=0.
\end{equation}
for $i,j,m,r=1,2,3,4$ with $i<j<m$, which becomes
\begin{equation}\label{eBo3HC4HS}
\begin{split}
&(4a-b^2)k_i^2k_j^2k_m^2\Big[2k_i^2w_jw_ml_jl_m+2k_j^2w_iw_ml_il_m+2k_m^2w_iw_jl_il_j
\\&+2k_jk_mw_jw_ml_i^2
+2k_ik_mw_iw_ml_j^2+2k_ik_jw_iw_jl_m^2+2k_jk_mw_i^2l_jl_m
\\&+2k_ik_mw_j^2l_il_m +2k_ik_jw_m^2l_il_j
-2k_ik_mw_jw_ml_il_j-2k_ik_mw_iw_jl_jl_m
\\&-2k_ik_jw_jw_ml_il_m-2k_ik_jw_iw_ml_jl_m
-2k_jk_mw_iw_ml_il_j-2k_jk_mw_iw_jl_il_m
\\&-k_i^2w_m^2l_j^2-k_i^2w_j^2l_m^2-k_j^2w_i^2l_m^2
-k_j^2w_m^2l_i^2-k_m^2w_j^2l_i^2-k_m^2w_i^2l_j^2\Big]=0
\end{split}
\end{equation}
where $i,j,m=1,2,3,4$, $i<j<m$. Some of the cases except $a=b^2/4$
which make this condition holds are;\\

\noindent \textbf{Case 1.} Any two of $k_i=0$, $i=1,2,3,4$, the
rest are
different,\\

\noindent \textbf{Case 2.} $k_i=\omega_i$, $i=1,2,3,4$,\\

\noindent \textbf{Case 3.} $k_i=l_i$, $i=1,2,3,4$,\\

\noindent \textbf{Case 4.} $\omega_i=l_i$, $i=1,2,3,4$.\\

\noindent The equation remaining from the coefficient of
$\varepsilon^4$ is
\begin{multline}
-P(\partial)f_4=e^{\theta_1+\theta_2+\theta_3+\theta_4}[B_{123}P(p_4-p_1-p_2-p_3)+B_{124}P(p_3-p_1-p_2-p_4)]\\
+B_{134}P(p_2-p_1-p_3-p_4)+B_{234}P(p_1-p_2-p_3-p_4)+A(1,2)A(3,4)P(p_1+p_2-p_3-p_4)
\\+A(1,3)A(2,4)P(p_1+p_3-p_2-p_4)+A(1,4)A(2,3)P(p_1+p_4-p_2-p_3)]=0.
\end{multline}
Hence $f_4=Ce^{\theta_1+\theta_2+\theta_3+\theta_4}$ where $C$ is
obtained as
\begin{equation}\label{eBCuzun}
\begin{split}
C=-[&A(1,2)A(3,4)P(p_1+p_2-p_3-p_4)
+A(1,3)A(2,4)P(p_1+p_3-p_2-p_4)\\+&A(1,4)A(2,3)P(p_1+p_4-p_2-p_3)
+B(1,2,3)P(p_4-p_1-p_2-p_3)\\+&B(1,2,4)P(p_3-p_1-p_2-p_4)
+B(1,3,4)P(p_2-p_1-p_3-p_4)\\+&B(2,3,4)P(p_1-p_2-p_3-p_4) ]\Bigg
/P(p_1+p_2+p_3+p_4).
\end{split}
\end{equation}
\noindent From the coefficient of $\varepsilon^5$ we have
\begin{equation}
2P(\partial)f_4+2P(D)\{f_1.f_3 \}+P(D)\{f_2.f_2 \}=0.
\end{equation}
\noindent The simplifications give us that
\begin{equation}\label{eBCkisa}
C=A(1,2)A(1,3)A(1,4)A(2,3)A(2,4)A(3,4).
\end{equation}
To be consistent the equations (\ref{eBCuzun}) and (\ref{eBCkisa})
should be equal to each other. This yields the four-Hirota
solution condition $(4HC)$
\begin{equation}
\sum_{\sigma_i=\pm 1}P(\sum_{i=1}^4 \sigma_i p_i) \prod_{0<i<j<4}
[P(\sigma_i p_i-\sigma_j p_j)]=0.
\end{equation}
In the hand, we know a case which satisfies both $(3HC)$ and
$(4HC)$ automatically which is\\

\noindent \textbf{Case 1.} Any two of $k_i=0$, $i=1,2,3,4$, the
rest are different.\\

\noindent Here we give the graphs of the two- and four-Hirota
solutions of eBo. We give arbitrary values to $a$, $b$, $k_i$ and
$w_i$. From the dispersion relation, we obtain $l_i$. We use these
constants in the solutions and draw their graphs.\\

\noindent \textbf{i)} \textbf{$N=2$, The Two-Hirota Solution of EKP}:\\

\noindent The constants are\\

\noindent $a=5$, $b=8$, $k_1=1$, $k_2=2$,\\

\noindent $w_1=-3$, $w_2=-5$, $\displaystyle
l_1=\frac{12+\sqrt{109}}{5}$, $\displaystyle
l_2=4-\sqrt{15}$.\\

\begin{center}
\begin{figure}[h]
\centering
\begin{minipage}[t]{0.4\linewidth}
\setlength{\fboxsep}{-\fboxrule}
\fbox{\includegraphics[angle=270,scale=.28]{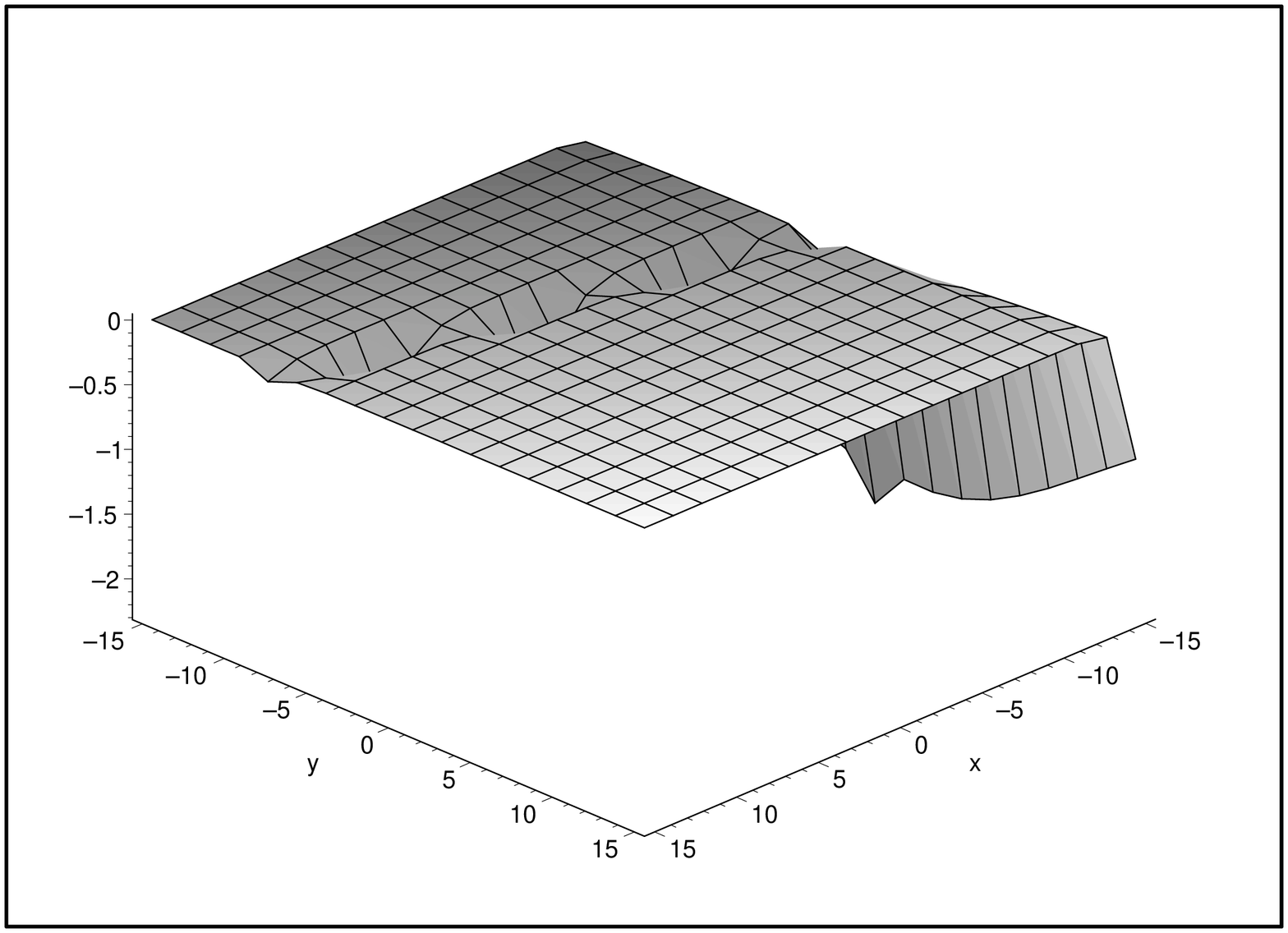}}
\caption{t=-6}
\end{minipage}%
\hspace{1cm}%
\begin{minipage}[t]{0.4\linewidth}
\setlength{\fboxsep}{-\fboxrule}
\fbox{\includegraphics[angle=270,scale=.28]{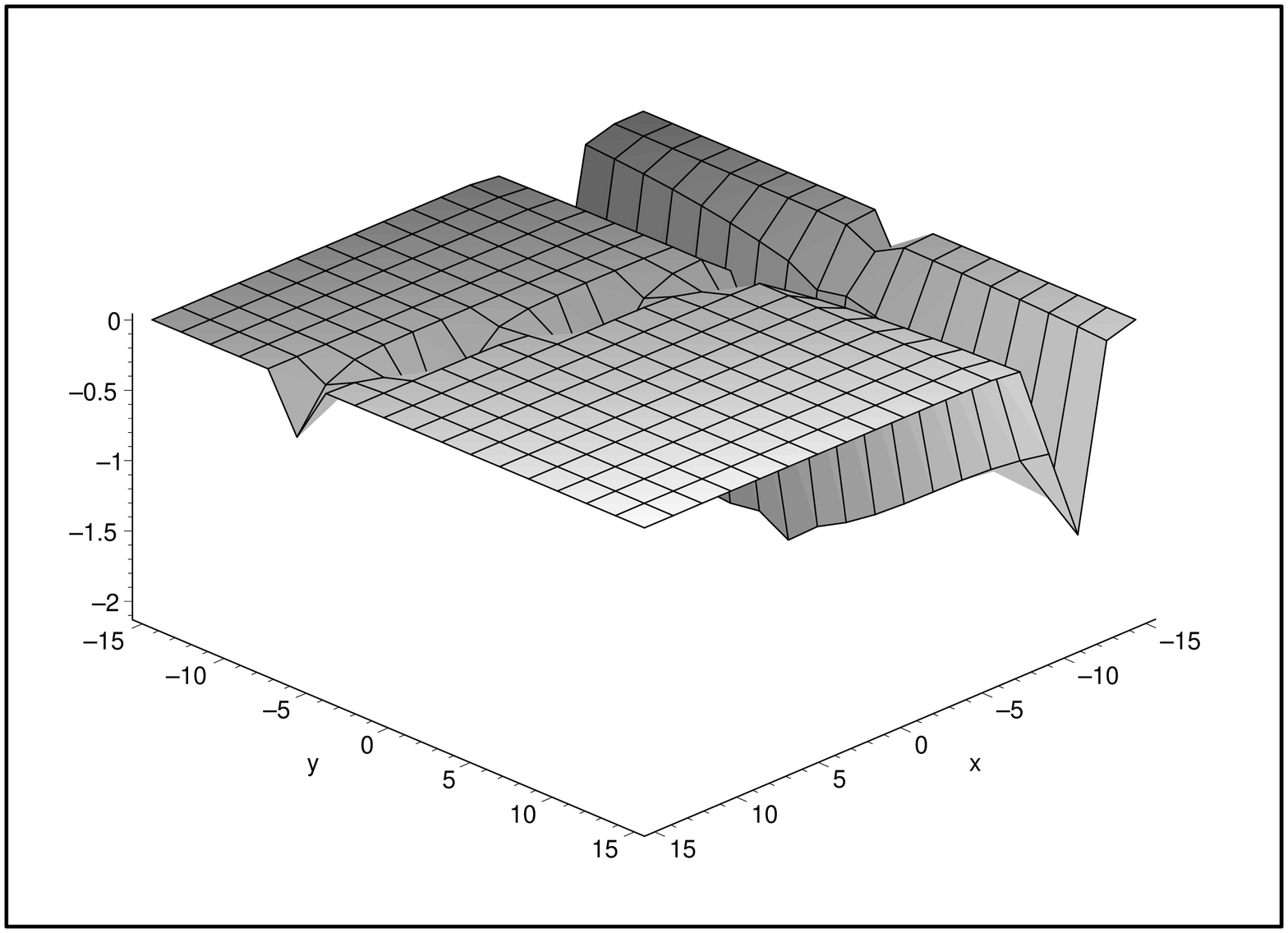}}
\caption{t=-4}
\end{minipage}
\end{figure}
\begin{figure}[h]
\centering
\begin{minipage}[t]{0.4\linewidth}
\setlength{\fboxsep}{-\fboxrule}
\fbox{\includegraphics[angle=270,scale=.28]{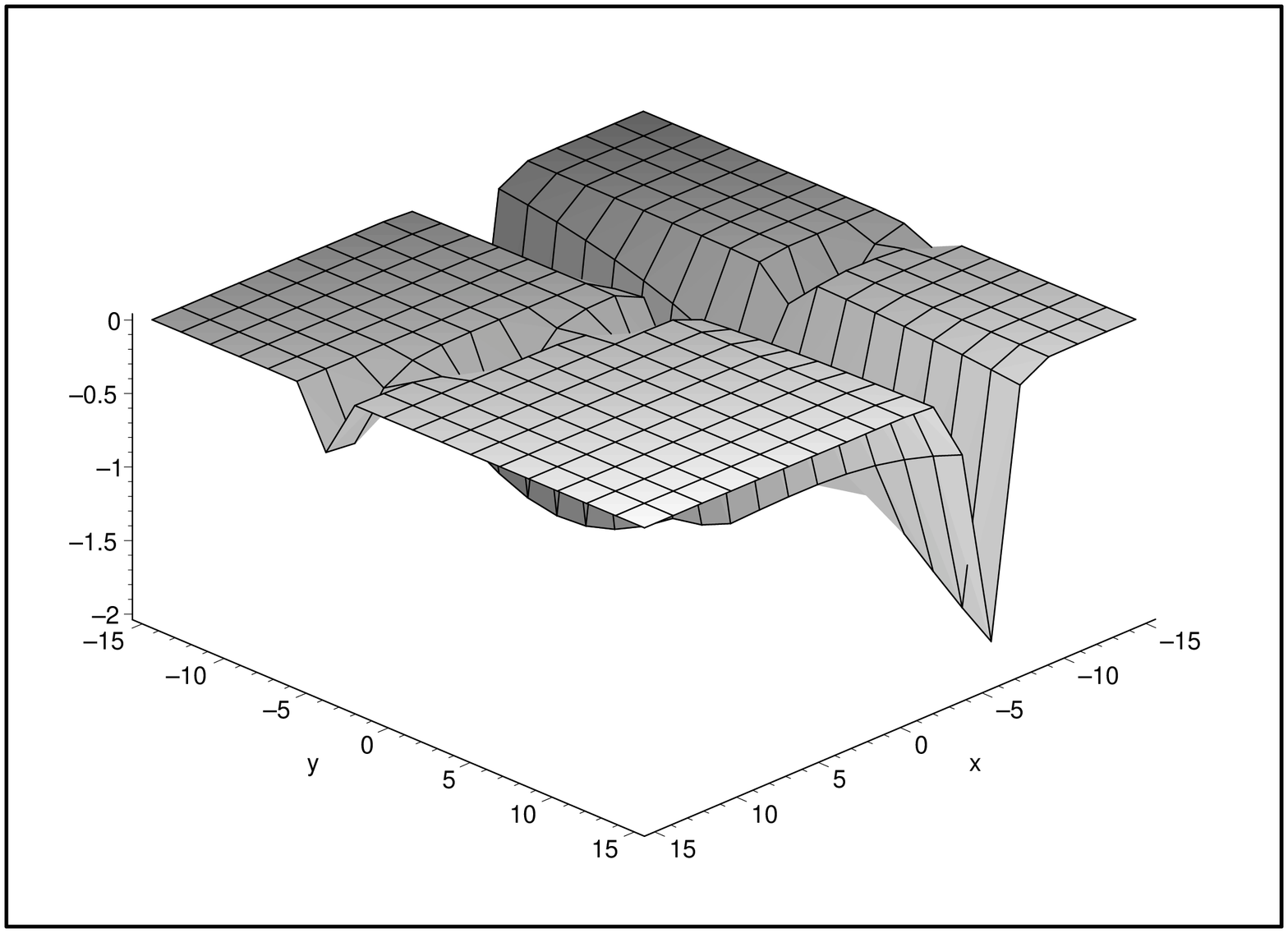}}
\caption{t=-2}
\end{minipage}%
\hspace{1cm}%
\begin{minipage}[t]{0.4\linewidth}
\setlength{\fboxsep}{-\fboxrule}
\fbox{\includegraphics[angle=270,scale=.28]{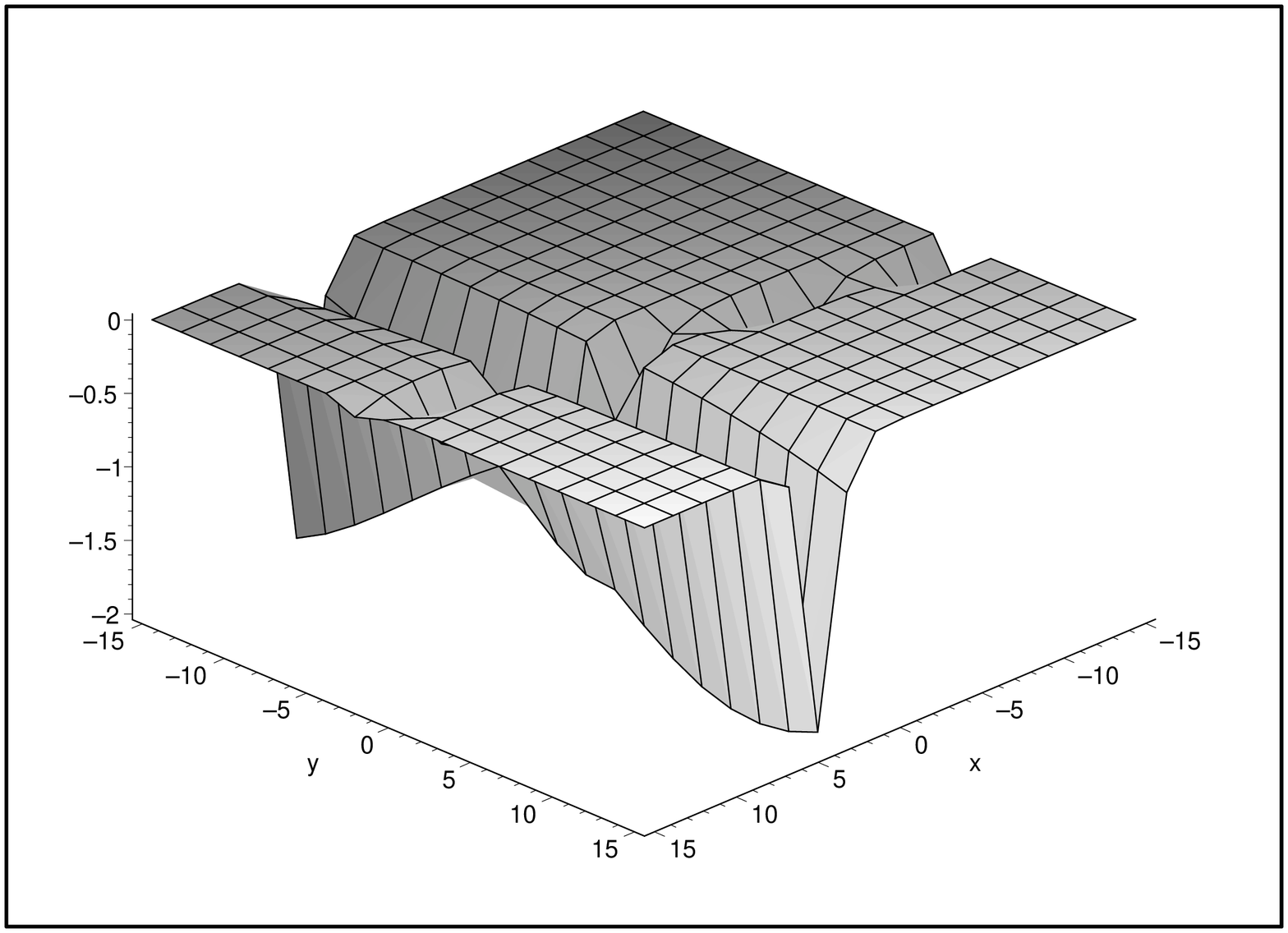}}
\caption{t=2}
\end{minipage}
\end{figure}
\newpage

\begin{figure}[h]
\centering
\begin{minipage}[t]{0.4\linewidth}
\setlength{\fboxsep}{-\fboxrule}
\fbox{\includegraphics[angle=270,scale=.28]{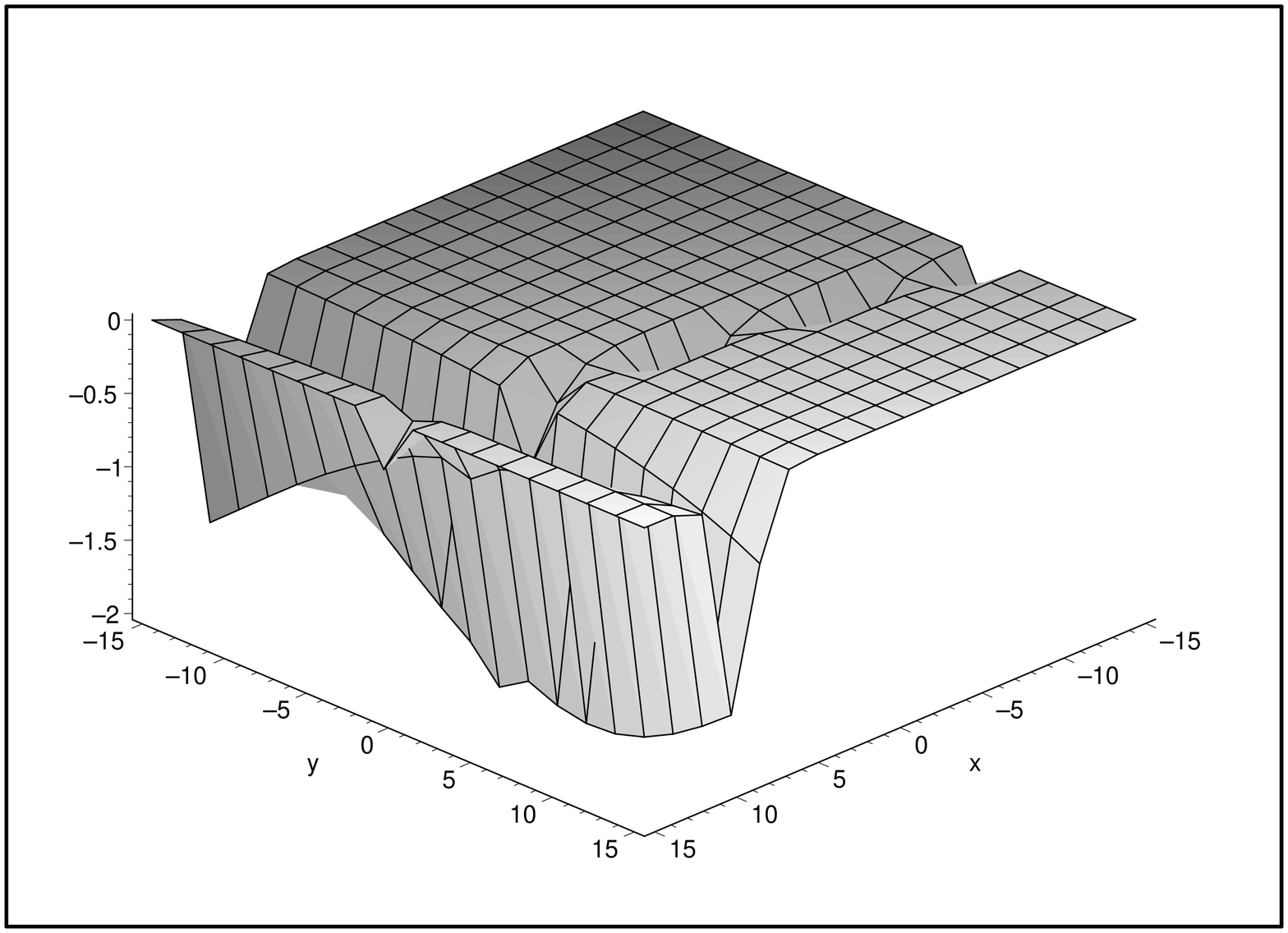}}
\caption{t=4}
\end{minipage}%
\hspace{1cm}%
\begin{minipage}[t]{0.4\linewidth}
\setlength{\fboxsep}{-\fboxrule}
\fbox{\includegraphics[angle=270,scale=.28]{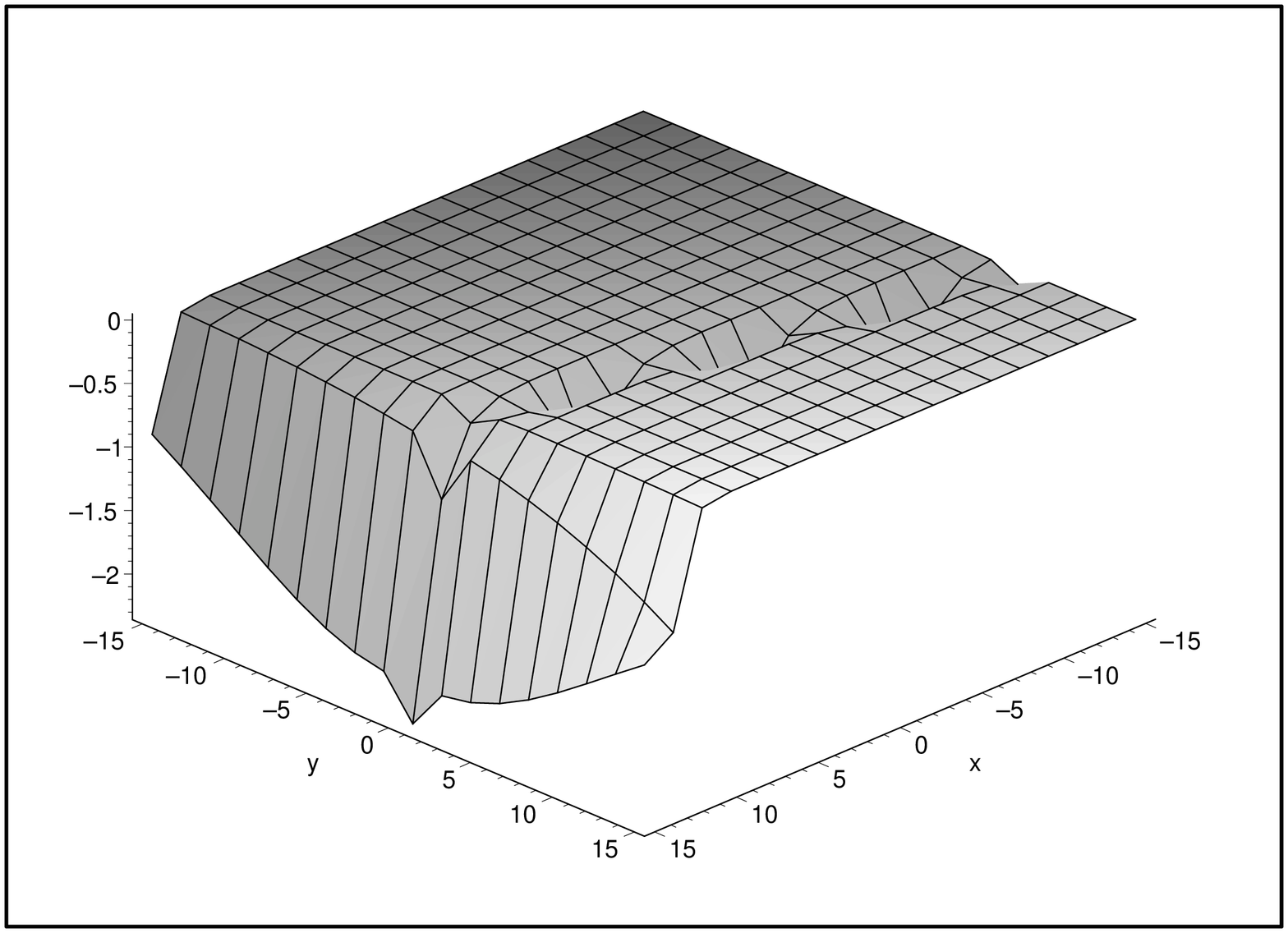}}
\caption{t=6}
\end{minipage}
\end{figure}
\end{center}

\noindent \textbf{ii)} \textbf{$N=4$, The Four-Hirota Solution of EBo}:\\

\noindent The constants are chosen according to the \textbf{Case
1} and the dispersion relation. The constants are,\\

\noindent $a=5$, $b=8$, $k_1=0$, $k_2=0$, $k_3=2$, $k_4=3$,\\

\noindent $w_1=5$, $w_2=4$, $w_3=-6$, $w_4=1$,\\

\noindent $\displaystyle l_1=-4+\sqrt{11}$, $\displaystyle
l_2=\frac{-16-4\sqrt{11}}{5}$, $\displaystyle
l_3=\frac{24+4\sqrt{31}}{5}$, $\displaystyle
l_4=\frac{-4-\sqrt{461}}{5}$.

\begin{center}
\begin{figure}[h]
\centering
\begin{minipage}[t]{0.4\linewidth}
\setlength{\fboxsep}{-\fboxrule}
\fbox{\includegraphics[angle=270,scale=.28]{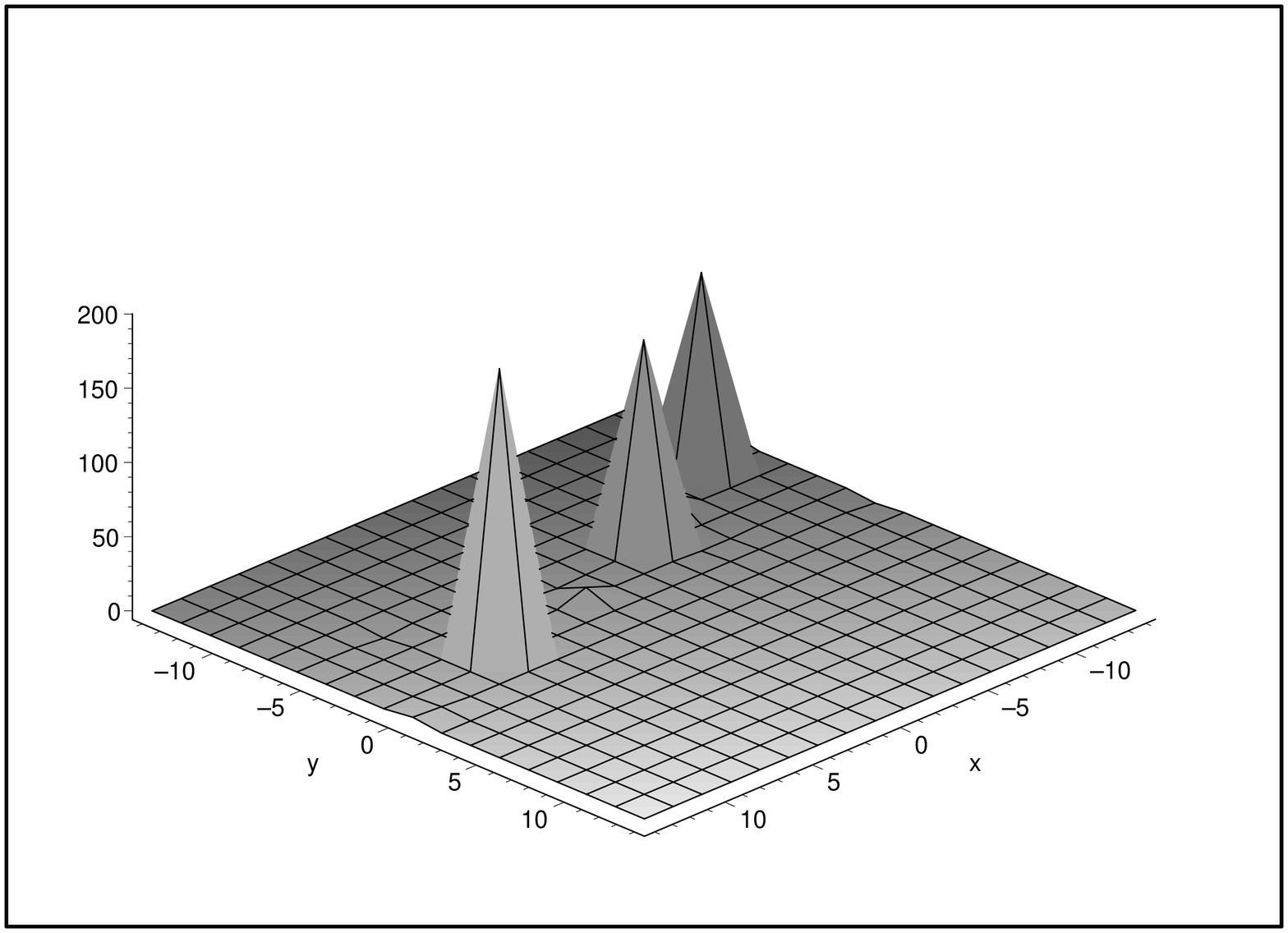}}
\caption{t=-6}
\end{minipage}%
\hspace{1cm}%
\begin{minipage}[t]{0.4\linewidth}
\setlength{\fboxsep}{-\fboxrule}
\fbox{\includegraphics[angle=270,scale=.28]{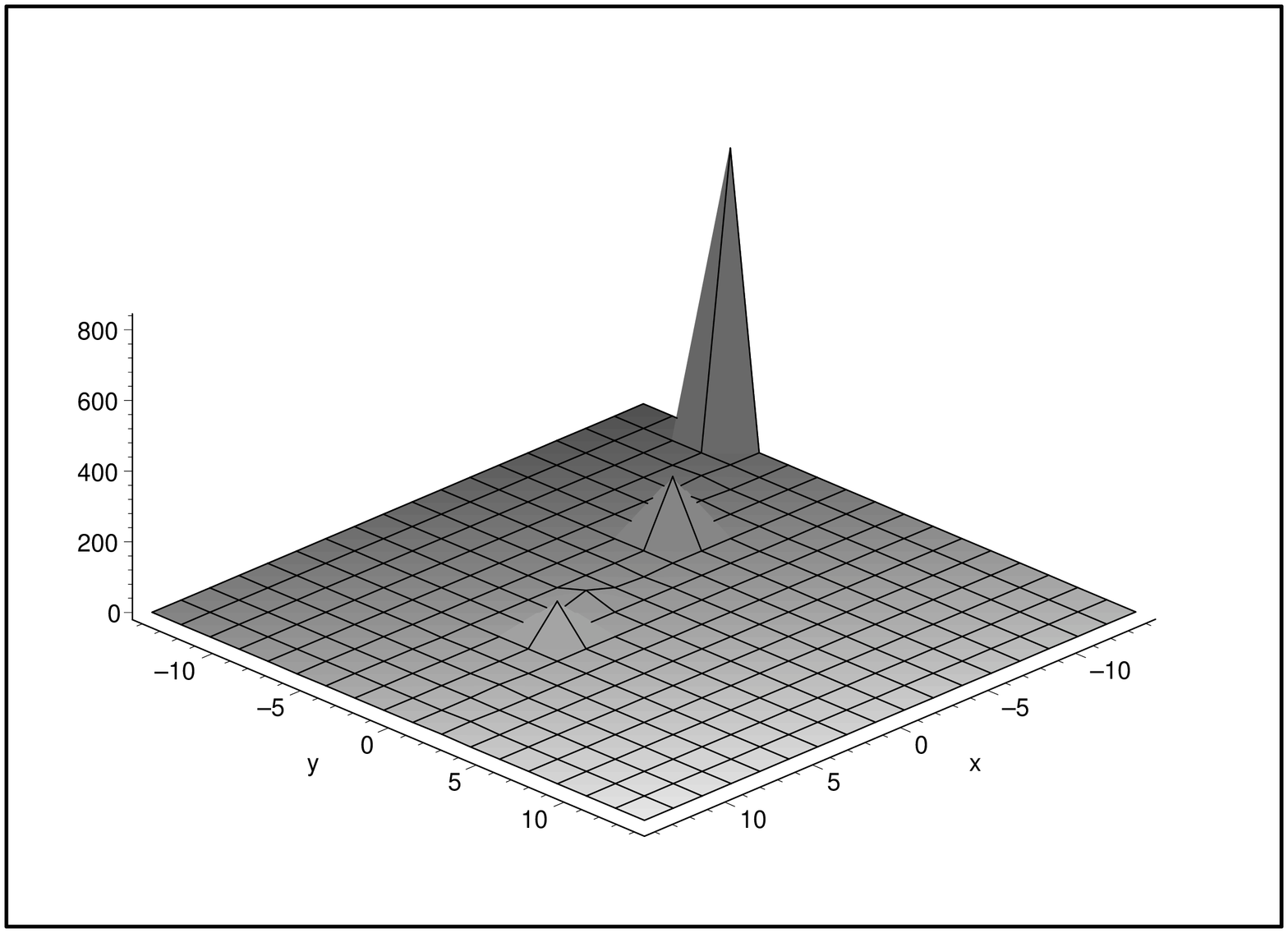}}
\caption{t=-4}
\end{minipage}
\end{figure}

\newpage

\begin{figure}[h]
\centering
\begin{minipage}[t]{0.4\linewidth}
\setlength{\fboxsep}{-\fboxrule}
\fbox{\includegraphics[angle=270,scale=.28]{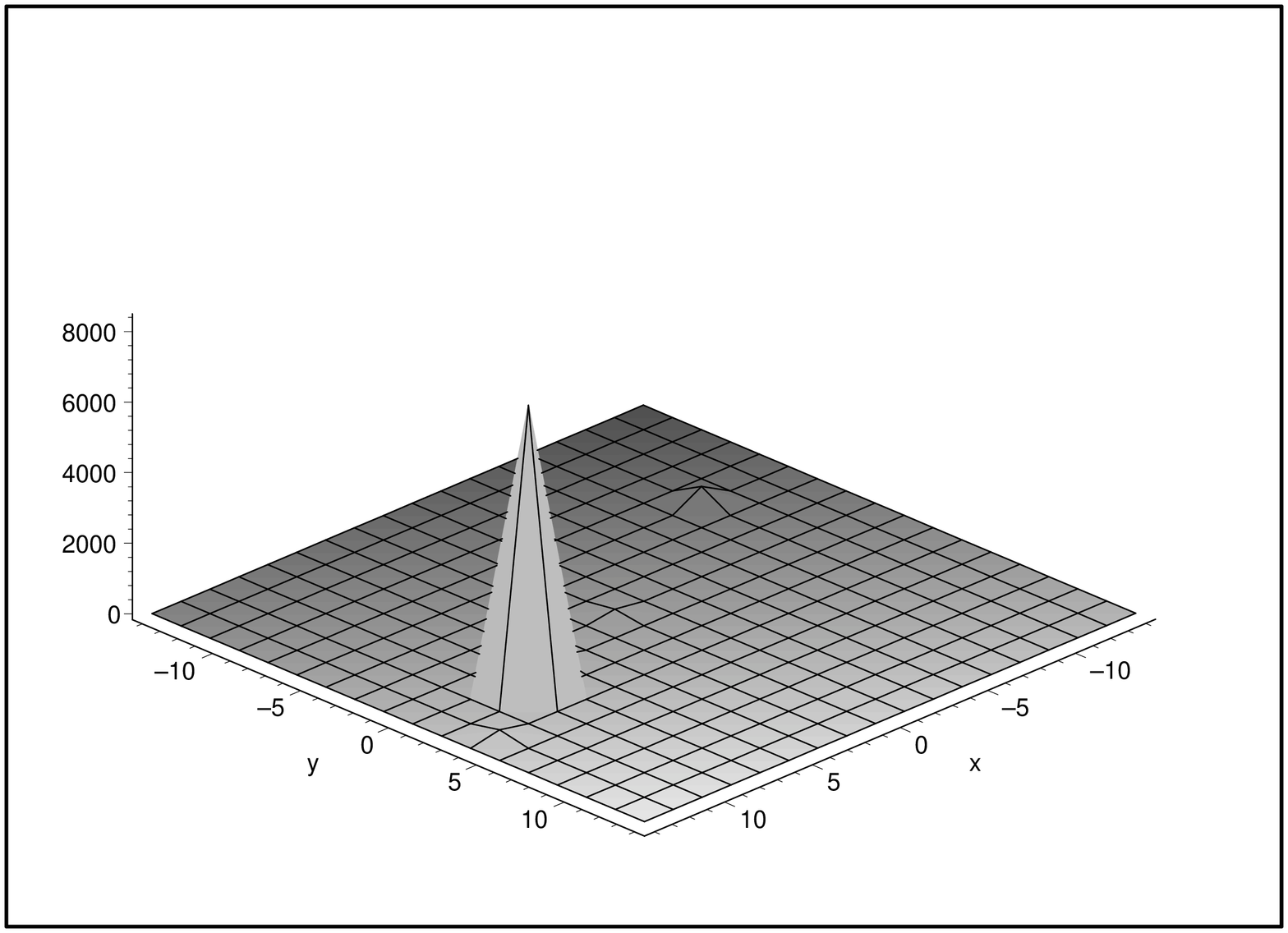}}
\caption{t=-2}
\end{minipage}%
\hspace{1cm}%
\begin{minipage}[t]{0.4\linewidth}
\setlength{\fboxsep}{-\fboxrule}
\fbox{\includegraphics[angle=270,scale=.28]{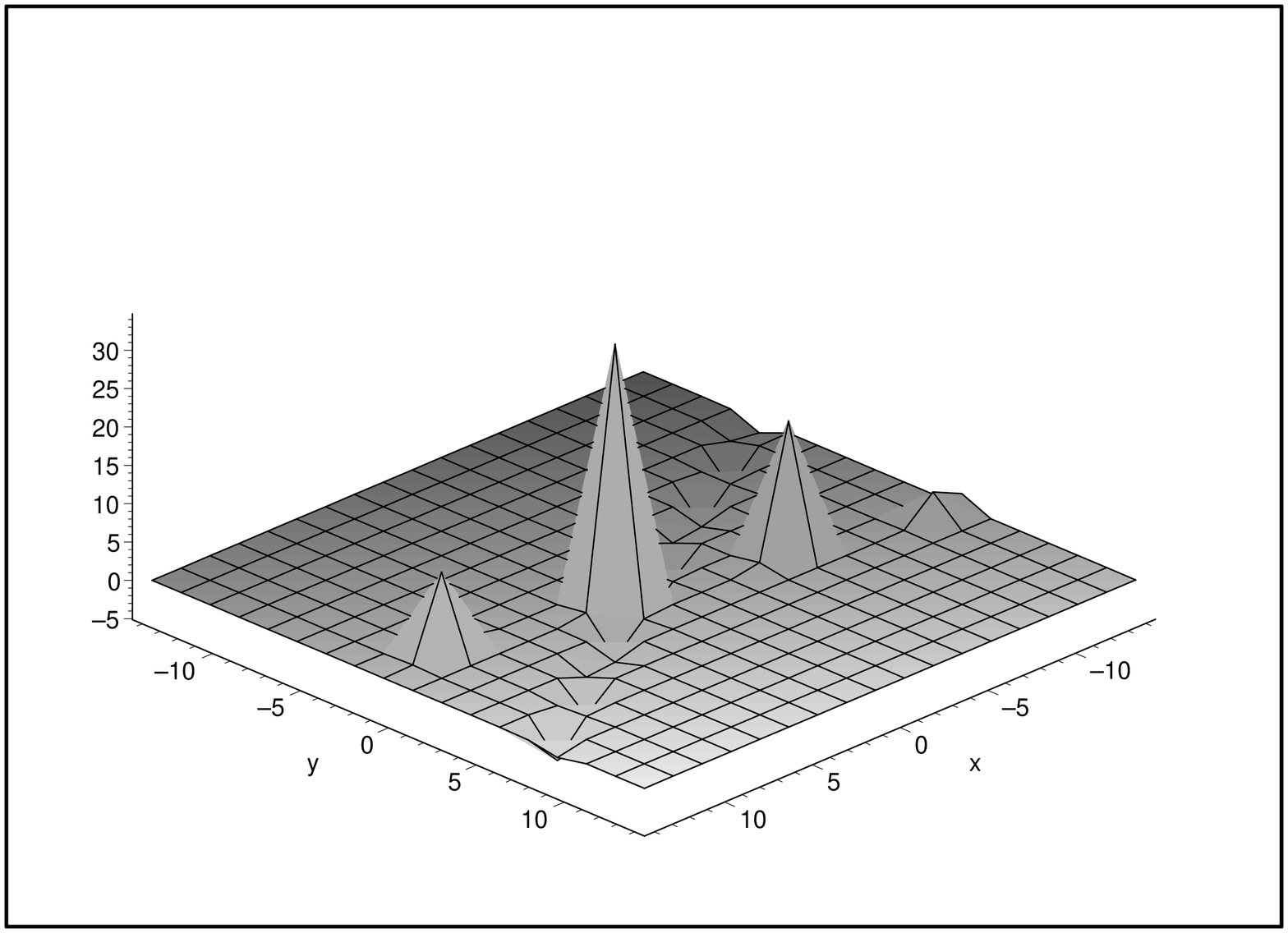}}
\caption{t=2}
\end{minipage}
\end{figure}
\begin{figure}[h]
\centering
\begin{minipage}[t]{0.4\linewidth}
\setlength{\fboxsep}{-\fboxrule}
\fbox{\includegraphics[angle=270,scale=.28]{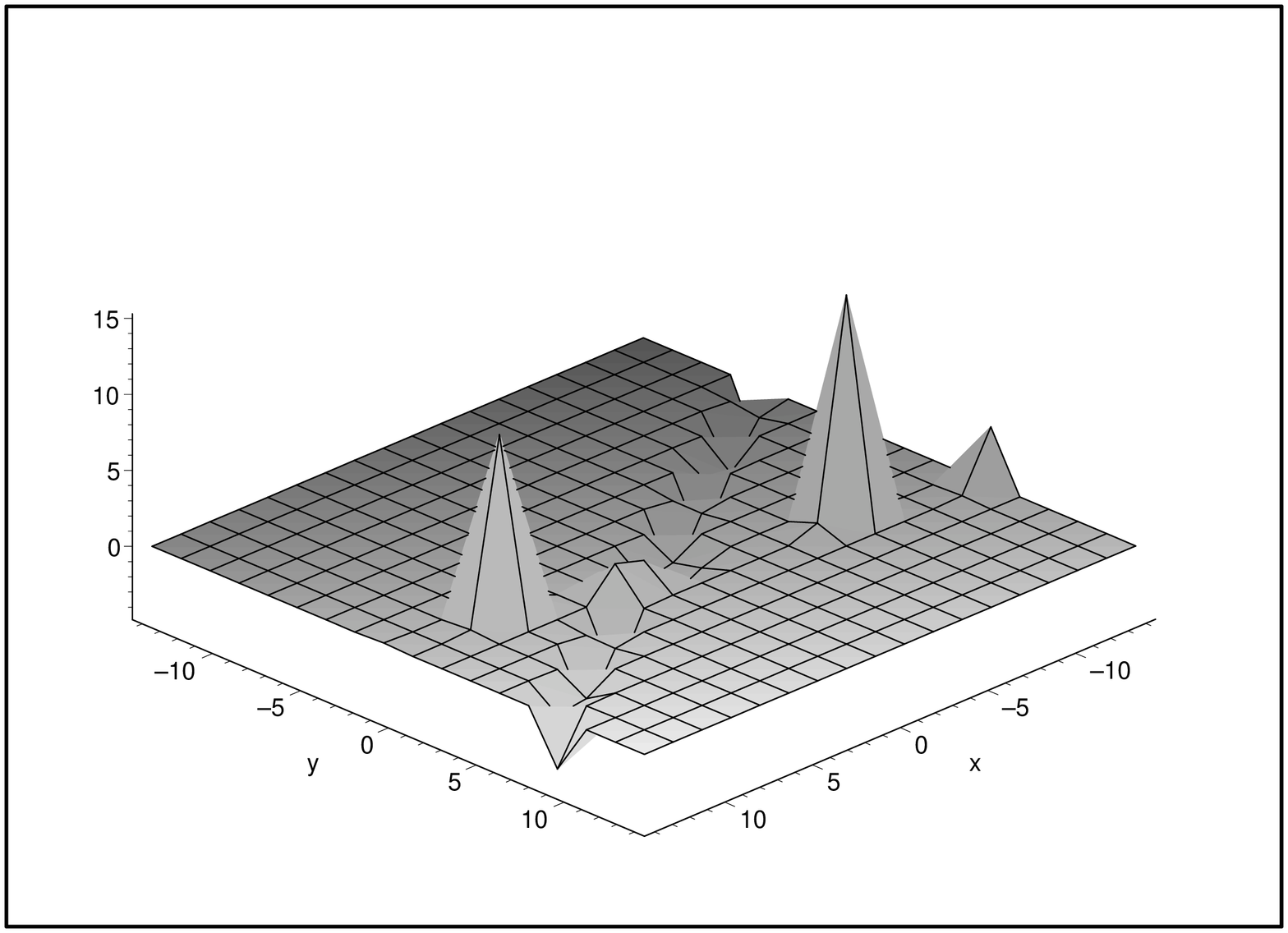}}
\caption{t=4}
\end{minipage}%
\hspace{1cm}%
\begin{minipage}[t]{0.4\linewidth}
\setlength{\fboxsep}{-\fboxrule}
\fbox{\includegraphics[angle=270,scale=.28]{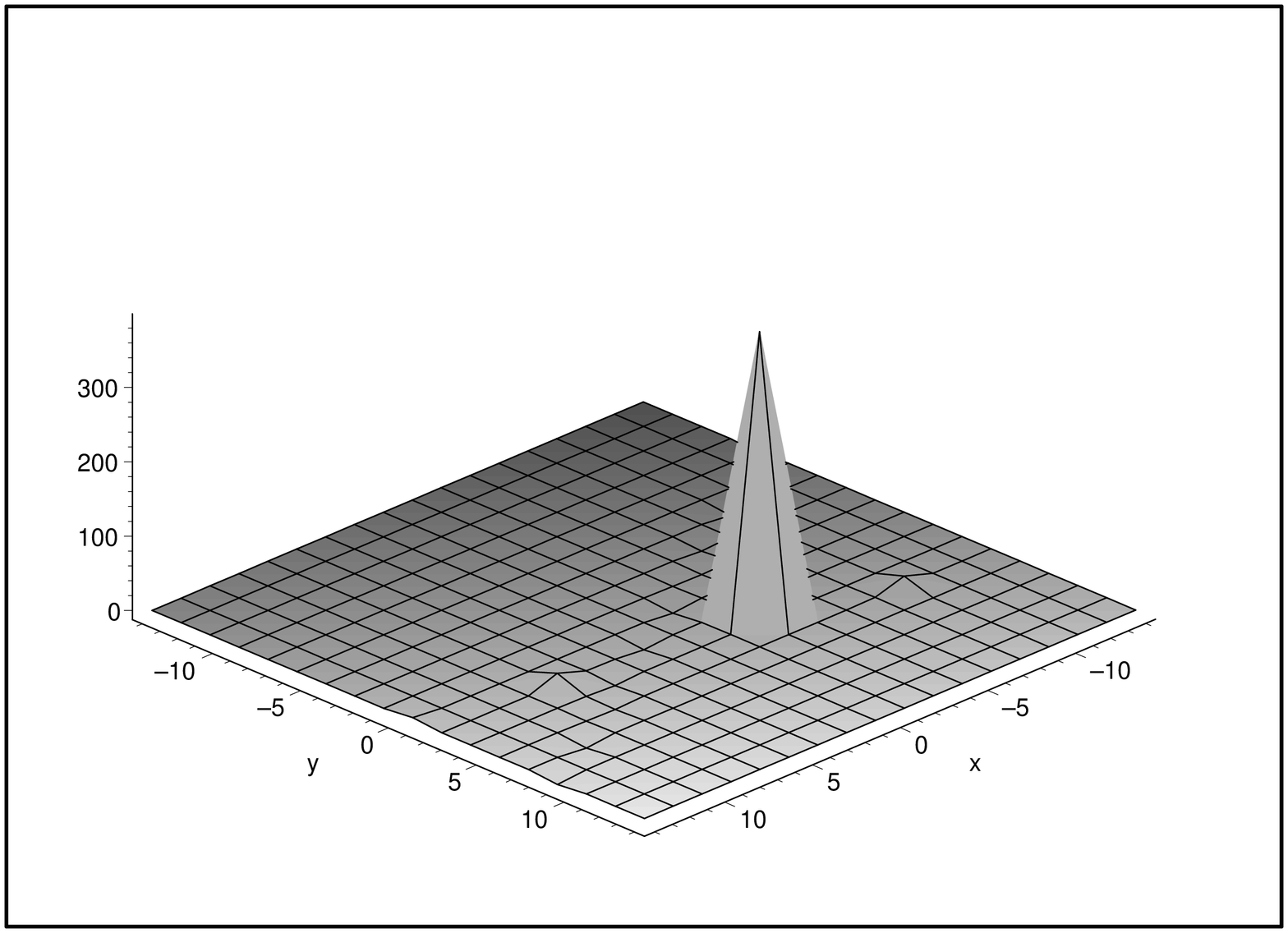}}
\caption{t=6}
\end{minipage}
\end{figure}
\end{center}

\vspace{12mm}

\section{Conclusion}
In this work, we applied the Hirota direct method to
non-integrable equations. We have given two examples, the extended
Kadomtsev-Petviashvili (eKP) and the extended Boussinesq
equations. They are in general non-integrable equations.

We have written bilinear and Hirota bilinear forms of these
equations. Since the equations having Hirota bilinear forms
automatically possess one- and two-Hirota solutions ( which have
soliton-like behavior), we have focused on three- and four-Hirota
solutions. We have seen that both equations should satisfy a
condition which we call three-Hirota solution condition $(3HC)$ to
have three-Hirota solution. While trying to obtain four-Hirota
solutions of the equations we have come across another condition,
four-Hirota solution condition $(4HC)$. We have classes of
solutions of these two conditions.

\section*{Acknowledgements}
This work is partially supported by the Scientific and Technical
Research Council of Turkey and Turkish Academy of Sciences.

\end{document}